\documentclass[twocolumn]{aastex6}

\usepackage{amsmath}
\usepackage{gensymb}
\usepackage{chngcntr}
\usepackage{longtable}

\fullcollaborationName{The Friends of AASTeX Collaboration}

\begin{document}


\title{Astrochemical Properties of Planck Cold Clumps}


\author{Ken'ichi Tatematsu\altaffilmark{1,2},
Tie Liu\altaffilmark{3},
Satoshi Ohashi\altaffilmark{4},
Patricio Sanhueza\altaffilmark{1},
Quang Nguy$\tilde{\hat{e}}$n Lu'o'ng\altaffilmark{1,3},
Tomoya Hirota\altaffilmark{1,2},
Sheng-Yuan Liu\altaffilmark{5},
Naomi Hirano\altaffilmark{5},
Minho Choi\altaffilmark{3},
Miju Kang\altaffilmark{3},
Mark Thompson\altaffilmark{6},
Garry Fuller\altaffilmark{7},
Yuefang Wu\altaffilmark{8},
Di Li\altaffilmark{9},
James di Francesco\altaffilmark{10,11},
Kee-Tae Kim\altaffilmark{3},
Ke Wang\altaffilmark{12},
Isabelle Ristorcelli\altaffilmark{13},
Mika Juvela\altaffilmark{14},
Hiroko Shinnaga\altaffilmark{15},
Maria Cunningham\altaffilmark{16},
Masao Saito\altaffilmark{17},
Jeong-Eun Lee\altaffilmark{18},
L. Viktor T\'oth\altaffilmark{19},
Jinhua He\altaffilmark{20,21,22},
Takeshi Sakai\altaffilmark{23},
Jungha Kim\altaffilmark{18},
JCMT Large Program ``SCOPE'' collaboration,
and
TRAO Key Science Program ``TOP'' collaboration
}


\altaffiltext{1}{National Astronomical Observatory of Japan,
National Institutes of Natural Sciences,
2-21-1 Osawa, Mitaka, Tokyo 181-8588, Japan; k.tatematsu@nao.ac.jp
}
\altaffiltext{2}{Department of Astronomical Science,
SOKENDAI (The Graduate University for Advanced Studies),
2-21-1 Osawa, Mitaka, Tokyo 181-8588, Japan}
\altaffiltext{3}{Korea Astronomy and Space Science Institute,
Daedeokdaero 776, Yuseong, Daejeon 305-348, South
Korea}
\altaffiltext{4}{Department of Astronomy, The University of Tokyo, Bunkyo-ku, Tokyo 113-0033, Japan}
\altaffiltext{5}{Academia Sinica Institute of Astronomy and Astrophysics, 11F of Astronomy-Mathematics Building, AS/NTU. No.1, Sec. 4, Roosevelt Rd, Taipei 10617, Taiwan, R.O.C.}
\altaffiltext{6}{Centre for Astrophysics Research, Science \& Technology Research Institute, University of Hertfordshire, Hatfield, AL10 9AB, UK}
\altaffiltext{7}{Jodrell Bank Centre for Astrophysics, School of Physics and Astronomy, University of Manchester, Oxford Road, Manchester, M13 9PL, UK}
\altaffiltext{8}{Department of Astronomy, Peking University, 100871, Beijing, China}
\altaffiltext{9}{National Astronomical Observatories, Chinese Academy of Sciences, Beijing, 100012, China}
\altaffiltext{10}{NRC Herzberg Astronomy and Astrophysics, 5071 West Saanich Rd, Victoria, BC V9E 2E7, Canada}
\altaffiltext{11}{Department of Physics and Astronomy, University of Victoria, Victoria, BC V8P 1A1, Canada}
\altaffiltext{12}{European Southern Observatory, Germany}
\altaffiltext{13}{IRAP, CNRS (UMR5277), Universit´e Paul Sabatier, 9 avenue du Colonel Roche, BP 44346, 31028, Toulouse Cedex 4, France}
\altaffiltext{14}{Department of physics, University of Helsinki, FI-00014, Helsinki, Finland}
\altaffiltext{15}{Department of Physics, Kagoshima University, 1-21-35, Korimoto, Kagoshima, 890-0065, Japan}
\altaffiltext{16}{School of Physics, University of New South Wales, Sydney, NSW 2052, Australia}
\altaffiltext{17}{Nobeyama Radio Observatory, National Astronomical Observatory of Japan, 
National Institutes of Natural Sciences, 
Nobeyama, Minamimaki, Minamisaku, Nagano 384-1305, Japan}
\altaffiltext{18}{School of Space Research, Kyung Hee University, Seocheon-Dong, Giheung-Gu, Yongin-Si, Gyeonggi-Do, 446-701, South Korea}
\altaffiltext{19}{Department of Astronomy, E\"otv\"os Lor\'and Unviersity, P\'azm´any P\'eter s\'etny 1, 1117 Budapest, Hungary}
\altaffiltext{20}{Key Laboratory for the Structure and Evolution of Celestial Objects, Yunnan Observatories, Chinese Academy of Sciences, P.O. Box 110, Kunming, 650011, Yunnan Province, China}
\altaffiltext{21}{Chinese Academy of Sciences, South America Center for Astrophysics (CASSACA), Camino El Observatorio 1515, Las Condes, Santiago, Chile}
\altaffiltext{22}{Departamento de Astronom\'ia, Universidad de Chile, Casilla 36-D, Santiago, Chile}
\altaffiltext{23}{Graduate School of Informatics and Engineering, The University of Electro-Communications, Chofu, Tokyo 182-8585, Japan}

\begin{abstract}
We observed thirteen Planck cold clumps with the James Clerk Maxwell Telescope/SCUBA-2 and with the Nobeyama 45 m radio telescope.  
The N$_2$H$^+$ distribution obtained with the Nobeyama telescope is quite similar to SCUBA-2 dust distribution.  The 82 GHz HC$_3$N, 82 GHz CCS, and 94 GHz CCS emission are often distributed  differently with respect to the N$_2$H$^+$ emission.  The CCS emission, which is known to be abundant in starless molecular cloud cores, is often very clumpy in the observed targets.  
We made deep single-pointing observations in DNC, HN$^{13}$C, N$_2$D$^+$, cyclic-C$_3$H$_2$
toward nine clumps.  The detection rate of N$_2$D$^+$ is 50\%.  
Furthermore, we observed the NH$_3$ emission toward 15 Planck cold clumps to estimate the kinetic temperature, and confirmed that most of targets are cold ($\lesssim$ 20 K).
In two of the starless clumps observe, the CCS emission is distributed as it surrounds the
N$_2$H$^+$ core (chemically evolved gas), which resembles the case of L1544, a prestellar core showing collapse. 
In addition, we detected both DNC and  N$_2$D$^+$.  These two clumps are most likely on the verge of star formation. 
We introduce the Chemical Evolution Factor (CEF) for starless cores to describe the chemical evolutionary stage, and analyze the observed Planck cold clumps.
\end{abstract}




\section{Introduction} \label{sec:intro}

On the basis of the Planck all-sky survey \citep{2011AA...536A..23P,2016A&A...594A..28P}, we are carrying out a series
of observations of molecular clouds as the Planck Cold Clump collaboration in order to
understand the initial condition for star formation \citep{2015PKAS...30...79L}. Planck cold clumps have low dust
temperatures (10$-$20 K; median=14.5 K). Pilot observations have been carried out with various ground-based telescopes such as
JCMT, IRAM, PMO 14m, APEX, Mopra, Effelsberg, CSO, and SMA \citep{2015PKAS...30...79L}.
A Large Program for JCMT dust continuum observations with SCUBA-2 (SCOPE
\footnote{https://www.eaobservatory.org/jcmt/science/large-programs/scope/}:
SCUBA-2 Continuum Observations of Pre-protostellar Evolution; Thompson et al. 2017, in preparation)
and a Key Science Program with TRAO 14 m radio telescope (TOP
\footnote{http://radio.kasi.re.kr/trao/key\_science.php}:
TRAO Observations of Planck cold clumps; Liu et al. 2017, in preparation)
are ongoing. 

To characterize Planck cold clumps, it is essential to investigate their chemical and  physical properties
in detail. In particular, we try to make their evolutionary stages clear. The chemical
evolution of molecular clouds has been established to some extent, not only for nearby dark clouds
(e.g., \cite{1992ApJ...392..551S,1992ApJ...394..539H,1998ApJ...506..743B,2006ApJ...646..258H,2009ApJ...699..585H}, but also for giant
molecular clouds (GMCs; \citet{2010PASJ...62.1473T,2014PASJ...66...16T,2014PASJ...66..119O} for Orion A GMC; \citet{2016PASJ...68....3O} for Vela C GMC; 
\citet{2012ApJ...756...60S,2013ApJ...777..157H} for Infrared Dark Clouds). Carbon-chain molecules such as CCS and HC$_3$N tend to be abundant in starless molecular cloud cores, while N-bearing molecules such as NH$_3$ and N$_2$H$^+$ as well as c-C$_3$H$_2$ tend to be abundant in
star-forming molecular cloud cores.
However, N$_2$H$^+$ will be destroyed by evaporated CO in warm cores having $T_{dust} \gtrsim$ 25 K \citep{2004ApJ...617..360L}, and therefore $N$(N$_2$H$^+$)/$N$(CCS) may not be a good evolutionary tracer for $T_{dust} \gtrsim$ 25 K.
In a survey of nearby cold dark cloud cores, \citet{1992ApJ...392..551S} detected CCS 45 GHz in
55\% of the observed cores, and found that CCS is a tracer of the young molecular gas in starless cloud cores.
\citet{2010PASJ...62.1473T} detected CCS$J_N$ = $4_3-3_2$ at 45 GHz in 32\% of the cores observed  toward Orion A GMC.
\citet{2008ApJ...678.1049S} observed 55 massive clumps associated with Infrared Dark Clouds (IRDCs) in CCS $J_N$ = $4_3-3_2$ at 45 GHz using the Nobeyama 45 m telescope, and detected this line in none of them.
\citet{2015AJ....150..159D} observed nine IRDCs in CCS $J_N$ = $2_1-1_0$ at 24 GHz using VLA, and detected this line in all of them.

Furthermore, deuterium fractionation ratios are powerful evolutionary tracers
\citep{2006ApJ...646..258H,2012ApJ...747..140S}.
The deuterium fraction D/H in molecules in molecular clouds is larger than the terrestrial abundance ratio (1.15$\times10^{-4}$)
or the protosolar estimate (2$\times10^{-5}$) \citep{1981AA....93..189G}.
The deuterium fraction is enhanced in molecular cloud cores before star formation (prestellar phase), which is cold enough, and the ratio increases with prestellar evolution until star formation \citep{2005ApJ...619..379C,2006ApJ...646..258H}. 
However, it should be noted that the deuterium fraction decreases with increasing temperature \citep{1979ApJ...228..748S,1987IAUS..120..311W,1992AA...256..595S,2010PASJ...62.1473T}.
After star formation, the deuterium fraction decreases \citep{2009A&A...496..731E}.
$N$(DNC)/$N$(HNC) will decrease, which may in part be due to an increase in the kinetic temperature, but may be affected by a larger reaction timescale (several times 10$^4$ yr)
while $N$(N$_2$D$^+$)/$N$(N$_2$H$^+$) will reflect the current temperature more directly, because of a shorter reaction time scale (less than 100 years)  \citep{2012ApJ...747..140S}.

We investigate the evolutionary stages of Planck cold clumps using molecular
column density ratios.
For this purpose, by using the Nobeyama 45 m telescope, we observed 13 Planck cold clumps, for
which we have already obtained accurate positions from preliminary IRAM 30m observations in N$_2$H$^+$ and/or SCUBA-2 observations.
We selected sources whose C$^{18}$O $J$ = 1$-$0 linewidths are relatively narrow ($<$ 1.5 km s$^{-1}$) from the previous observations
with the 13.7 m telescope of the Purple Mountain Observatory (PMO) at De Ling Ha, because we prefer to observe relatively close objects
to investigate the chemical differentiation on scales of 0.05-0.1 pc
\citep{2014PASJ...66...16T,2014PASJ...66..119O,2016PASJ...68....3O}.
However, some of our sources (G108.8-00.8 and G120.7+2.7) are actually distant (1-3.5 kpc), which needs some care in discussion based on the column density ratios.
While we focus on the chemical evolution stage as the main target of the current study, we also briefly investigate the physical properties of the sources by analyzing the specific angular momenta.

\section{Observations}


\subsection{James Clerk Maxwell Telescope}

Observations with the 15 m James Clerk Maxwell Telescope (JCMT) on Mauna Kea were made between 2014 November and 2015 December in the pilot survey phase (project IDs: M15AI05, M15BI061) 
of the JCMT legacy survey program ``SCOPE".  The Submillimetre Common-User Bolometer Array 2 (SCUBA-2) was employed for observations of the 850 $\mu$m continuum. It is a 10,000 pixel bolometer camera operating simultaneously at 450 and 850 $\micron$.  Observations were carried out in constant velocity (CV) Daisy mode under grade 3/4 weather conditions with a 225 GHz opacity between 0.1 and 0.15. The mapping area is about $12\arcmin\times$12\arcmin. The beam size of SCUBA-2 at 850 $\micron$ is $\sim$14$\arcsec$. 
The typical rms noise level of the maps is about 6-10 mJy~beam$^{-1}$ in the central 3$\arcmin$ area, and increases to 10-30 mJy~beam$^{-1}$
out to 12$\arcmin$.  The data were reduced using SMURF in the STARLINK package.

\subsection{Nobeyama 45 m Telescope}

Observations with the 45 m radio telescope
of Nobeyama Radio Observatory\footnote{Nobeyama Radio Observatory
is a branch of the National Astronomical Observatory of Japan,
National Institutes of Natural Sciences.} were carried out from 2015 December to  2016 February.
We observed eight molecular lines by using the receivers TZ1 and T70 (Table 1).
Observations with the receiver TZ1 (we used one beam called TZ1 out of two beams of the receiver TZ) \citep{2013PASP..125..213A,2013PASP..125..252N} were made 
to simultaneously observe four molecular lines,
82 GHz CCS, 94 GHz CCS, HC$_3$N and N$_2$H$^+$.
Observations were carried out with T70 to simultaneously observe four other lines,
HN$^{13}$C, DNC, N$_2$D$^+$, and cyclic-C$_3$H$_2$.
TZ1 and T70 are double-polarization, two-sideband SIS receivers.
Molecular transitions
are selected to achieve a high angular resolution of $\sim$20" to differentiate the distribution of molecules, 
and they are
detectable even from cold gas ($\lesssim$ 20 K).
The upper energy level $E_u$ of the observed transitions are listed in Table 1.

The FWHM beam sizes at 86 GHz with TZ1 and T70 were 18$\farcs$2$\pm$0$\farcs$1 and 18$\farcs$8$\pm$0$\farcs$3, respectively.
The main-beam efficiency $\eta_{mb}$ at 86 GHz with TZ1 was 54$\pm$3\% and 53$\pm$3\% in H and V polarizations, respectively.
The efficiencies with T70 was 54$\pm$3\% and 55$\pm$3\% in H and V polarizations, respectively.
We also observed NH$_3$ $(J, K)$ = (1, 1) at 23.694495 GHz \citep{1983ARAA..21..239H}
in both circular polarizations
with the receiver H22.
The FWHM beam size at 23 GHz with H22 was 74$\farcs$4$\pm$0$\farcs$3 and 73$\farcs$9$\pm$0$\farcs$3 for ch1 and ch2 (right-handed and left-handed circular polarization), respectively.
The main-beam efficiency $\eta_{mb}$ at 23 GHz with H22 was 83$\pm$4\% and 84$\pm$4\%, for ch1 and ch2, respectively.
The receiver backend was the digital spectrometer ``SAM45''.
The spectral resolution
was 15.26 kHz  (corresponding to 0.05$-$0.06 km s$^{-1}$) for TZ1 and T70,
and 3.81 kHz (corresponding to $\sim$0.05 km s$^{-1}$) for H22.

Observations with receiver TZ1 were conducted in the on-the-fly (OTF) mapping mode \citep{2008PASJ...60..445S} 
with data sampling intervals along a strip of 5$\arcsec$ and separations between strips of 5$\arcsec$.
We have made orthogonal scans in RA and DEC to minimize scan effects.
Observations with receivers T70 and H22 were carried out in the ON-OFF position-switching mode.
The ON positions with receiver T70 were determined from possible intensity peaks in N$_2$H$^+$ on temporal lower-S/N-ratio TZ1 maps during observations (before completion), but are not necessarily intensity peaks in N$_2$H$^+$ on the final maps.
In addition, the T70 position of G207N was incorrect.

Coordinates used for the observations are listed in Table 2.
The distances to the sources are taken from
\citet{2011AA...535A...8R} for G108.8-00.8,
\citet{2012ApJ...756...76W} for G120.7+2.7,
\citet{2010AA...512A..67L} for G157.6-12.2,
\citet{2007ApJ...671..546L} and \citet{2010AA...512A..67L} for G174.0-15.8,
\citet{2008PASJ...60..991K} for G192.33-11.88, G202.31-8.92, G204.4-11.3, and G207.3-19.8,
and \citet{1974AJ.....79.1022C} for G224.4-0.6.
Those to G089.9-01.9,	
G149.5-1.2, and
G202.00+2.65
are taken from
\citet{2016A&A...594A..28P}.

The observed intensity is reported in terms of the corrected
antenna temperature $T_A^*$.
To derive the physical parameters, we use the main-beam
radiation temperature $T_{mb}$ = $T_A^*$/$\eta_{mb}$.
The telescope pointing was established by observing relevant 43-GHz SiO maser sources every $\sim$60 min for TZ1 and T70 observations,
and was accurate to $\sim$5$\arcsec$.  For H22 observations, pointing was established once a day in the beginning of observations.
The observed data were reduced using the software package ``NoStar'' and ``NewStar''
of Nobeyama Radio Observatory.




\section{Results}

Figures 1 through 13 show the maps obtained with JCMT/SCUBA-2 and with the Nobeyama 45 m telescope.
Table 3 lists the dense cores detected with JCMT/SCUBA-2, their intensities, and column densities.
The column density is derived by using equation (A.27) in \citet{2007PhDT.......435K}.
Associations of young stars are investigated on the basis of WISE 22$\mu$m, Akari, Spitzer images and IRAS point source catalog,
and they are classified 
through their spectral indexes (Liu et al. 2016, in preparation).
N$_2$H$^+$ was detected for all sources,
and we detected 81 GHz CCS from seven out of 13.
Tables 4 and 5 summarize observed parameters from Nobeyama 45 m telescope receiver TZ1.
The emission line spectra have a single velocity component, when they are detected,  and there is no position showing two velocity components.
For Table 5, we should note that
T70 positions do not necessarily coincide with the final N$_2$H$^+$ peaks as intended.
The line parameters are obtained through Gaussian fitting (either single-component Gaussian or hyperfine fitting).
The upper limit to the intensity is defined as 3$\sigma$, where $\sigma$ is the rms noise level at 1 km s$^{-1}$ bin.

Tables 6 summarizes the observed parameters from the Nobeyama 45 m telescope receiver T70.
Again, emission line spectra have single velocity component, when they are detected, and there is no position showing two velocity components.
We detected DNC, HN$^{13}$C, c-C$_3$H$_2$ all but G108N and G207N (again, G207N position was mistakenly set).
The N$_2$D$^+$ hyperfine transitions were detected toward four out of eight sources (excluding G207N).

Furthermore, we observed 15 sources with the Nobeyama 45 m telescope receiver H22 in NH$_3$ transitions
in single-pointing observations.
Nine out of them were detected in $(J,K) = (2,2)$, and we derived
$T_{rot}$. 
The NH$_3$ rotation temperature $T_{rot}$ is derived as explained in \citet{1983ARAA..21..239H},
and then converted to the gas kinetic temperature $T_k$ by using the relation by \citet{2004AA...416..191T}.
Observed intensities, derived $T_{rot}$ and $T_k$ are summarized in Table 7.

We investigate the molecular line intensity distribution shown in Figures 1 to 13.
In general, the N$_2$H$^+$ distribution (black contours in panel (b)) is quite similar to the 850 $\mu$m dust continuum emission distribution (contours in panel (a); gray-scale in panels (b) and (c)).
The 82 GHz CCS emission (blue contours) is clumpy in general, and is often located at the edge of the N$_2$H$^+$/850 $\mu$m  core or is distributed as it surrounds the N$_2$H$^+$/850 $\mu$m  core.
We draw the 2.5$\sigma$ contour in some cases to let the reader understand the reliablity of the 3$\sigma$ contour as detection.
Most clumps are as cold as 10$-$20 K, and therefore the depletion of CCS can contribute to a configuration of the N$_2$H$^+$ core being surrounded by CCS \citep{2001ApJ...552..639A,2002ApJ...570L.101B}.
The 94 GHz CCS emission, when detected, does not necessarily follow that of the 82 GHz CCS emission, although
the upper energy levels of these transitions do not differ so much.
Taking the beamsizes of JCMT/SCUBA-2 and the Nobeyama 45 m telescope and the pointing accuracy of the latter telescope, only 
differences in the spatial distribution larger than 10$\arcsec$
will be meaningful.

The molecular column density is calculated by assuming local thermodynamic equilibrium (LTE) as explained in, e.g., \citet{1992ApJ...392..551S, 2015PASP..127..266M,2012ApJ...756...60S}.  N$_2$H$^+$, N$_2$D$^+$, and NH$_3$ have hyperfine transitions, and we can derive the excitation temperatures $T_{ex}$ directly from observations.
We assume that the beam filling factor is unity.
If the filling factor is lower than unity, we underestimate $T_{ex}$.
$T_{ex}$ from N$_2$H$^+$ ranges from 3.7 to 5.7 K.
When we compared $T_{ex}$ for N$_2$H$^+$ with  $T_K$ derived from NH$_3$ observations, we obtain a relation of $T_{ex}$ = 0.45$\pm$0.27 $T_K$.  
We assume that the excitation temperatures $T_{ex}$ for the other molecular lines are 0.5 $T_K$, assuming that the levels are subthermally excited to 50\% (for G207N, we assume $T_{ex}$ = 0.6 $T_K$ so that the observed intensity can be explained in LTE).
If $T_k$ is an upper limit, we assume $T_k$ = 10 K.
The necessary parameters for CCS are taken from \citet{1990ApJ...361..318Y} and references therein.
For  N$_2$H$^+$, if the hyperfine fitting is unsuccessful, we try to calculate the column density by using the velocity-integrated intensity of  
a main  $F_1$ = 2$-$1 group of the three hyperfine components ($F$ =  1$-$0,  2$-$1, and 3$-$2), by assuming optically thin emission,
and by neglecting the background term.

We show in Table 8 the column density range by assuming that the actual column density is a factor of 1$-$10 larger than the optically thin estimate.
The resulting column densities are listed in Tables 8 and 9.
Although we observed the 94 GHz CCS emission, the detection rate is not high. At T70 positions, only two clumps were detected but the S/N ratio is low. 
We decided to use $T_k$ to estimate $T_{ex}$ (CCS) instead of the excitation analysis (e.g., Large Velocity Gradient models)  to treat all the data in a consistent way.
Tables 10 and 11 list the fractional abundance $X$ of molecules relative to H$_2$ calculated from the column density ratio toward the JCMT/SCUBA-2 peaks and T70 positions, respectively.  The H$_2$ column density is taken from the dust continuum flux density measured toward the T70 position.

In the next section, we introduce the results for individual sources.

\begin{figure*}
\includegraphics[angle=0,scale=0.6]{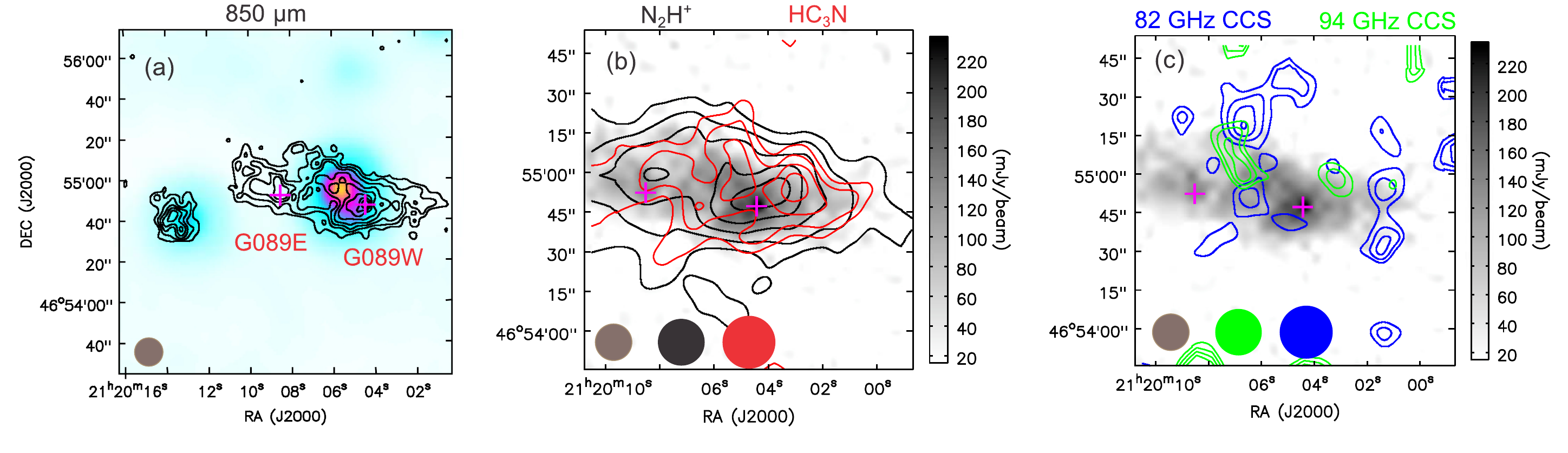}
\caption{(a) The 850 $\mu$m continuum image (contours) obtained toward G089.9-01.9 superimposed on the WISE 22 $\mu$m image (color).
Contour levels are
[0.2, 0.3, 0.4, 0.5, 0.6, 0.7, 0.8, 0.9] $\times$
235.7 mJy/beam.
The magenta cross symbol represents the SCUBA-2 peak position (Table 3).
The dark brown filled circle represents the half-power beam size for  JCMT/SCUBA-2 850 $\mu$m.
(b)
Black and red contour maps represent N$_2$H$^+$ and HC$_3$N integrated intensity maps, respectively, while
the gray-scale map represents the 850 $\mu$m continuum intensity.
The N$_2$H$^+$  integrated intensity is calculated for the main hyperfine component group $F_1$ = 2$-$1 ($F$ =  1$-$0,  2$-$1, and 3$-$2)
from a velocity range of
0.5 to 3.5 km s$^{-1}$
with levels of
[3, 6, 9, 12, 15] $\times$
0.092 K km s$^{-1}$(1$\sigma$).
The HC$_3$N intensity is calculated from a velocity range of
0.5 to 3.5 km s$^{-1}$
with levels of
[3, 4.5, 6, 7.5, 10] $\times$
0.081 K km s$^{-1}$(1$\sigma$).
The black and red filled circles represent the half-power beam sizes for  N$_2$H$^+$ and HC$_3$N, respectively.
(c) Blue and green contours represent the 82 GHZ CCS and 94 GHz CCS integrated intensity maps, respectively, while 
the gray-scale map represents the 850 $\mu$m continuum intensity.
The 82 GHz CCS intensity is calculated from a velocity range of
0.5 to 3.5 km s$^{-1}$
with levels of
[2.5, 3, 3.5, 4] $\times$
0.080 K km s$^{-1}$(1$\sigma$).
The 94 GHz CCS intensity is calculated from a velocity range of
0.5 to 3.5 km s$^{-1}$
with levels of
[2.5, 3, 3.5, 4] $\times$
0.11 K km s$^{-1}$(1$\sigma$).
The blue and green filled circles represent the half-power beam sizes for  82 GHz CCS and 94 GHz CCS, respectively.
}
\end{figure*}

\begin{figure*}
\includegraphics[angle=0,scale=0.4]{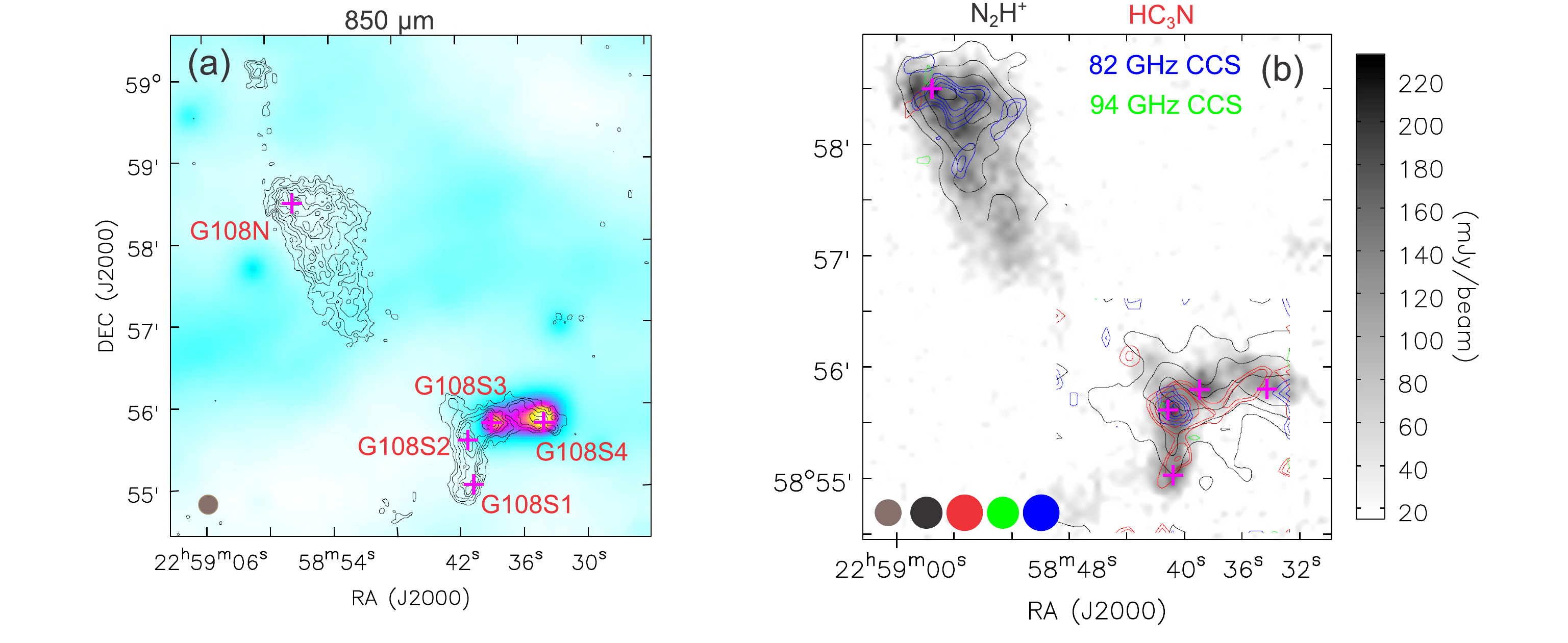}
\caption{Same as Figure 1 but for G108.8-00.8.
(a) The 850 $\mu$m continuum (contours) levels are
[0.2, 0.3, 0.4, 0.5, 0.6, 0.7, 0.8, 0.9] $\times$
231.9 mJy/beam.
(b) 
For G108N,
The N$_2$H$^+$ intensity (black contours) is calculated from a velocity range of
-51.5 to -47.5 km s$^{-1}$
with levels of
[3, 6, 9, 12, 15] $\times$
0.088 K km s$^{-1}$(1$\sigma$).
For G108S,
the N$_2$H$^+$ intensity (black contours) is calculated from a velocity range of
-52.5 to -48.5 km s$^{-1}$
with levels of
[3, 6, 9, 12, 15] $\times$
0.066 K km s$^{-1}$(1$\sigma$).
For G108N,
the HC$_3$N intensity (red) is calculated from a velocity range of
-51.5 to -47.5 km s$^{-1}$
with levels of
2.5 $\times$
0.078 K km s$^{-1}$(1$\sigma$).
For G108S,
the HC$_3$N intensity (red) is calculated from a velocity range of
-52.5 to -48.5 km s$^{-1}$
with levels of
[2.5, 3, 4, 5, 6] $\times$
0.064 K km s$^{-1}$(1$\sigma$).
For G108N,
the 82 GHz CCS intensity (blue) is calculated from a velocity range of
-51.5 to -47.5 km s$^{-1}$
with levels of
[2.5, 3, 3.5, 4, 4.5] $\times$
0.080 K km s$^{-1}$(1$\sigma$).
For G108S,
the 82 GHz CCS intensity (blue) is calculated from a velocity range of
-52.5 to -48.5 km s$^{-1}$
with levels of
[2.5, 3, 3.5, 4, 4.5] $\times$
0.078 K km s$^{-1}$(1$\sigma$).
For G108N,
the 94 GHz CCS intensity (green) is calculated from a velocity range of
-51.5 to -47.5 km s$^{-1}$
with levels of
2.5 $\times$
0.11 K km s$^{-1}$(1$\sigma$).
For G108S,
the 94 GHz CCS intensity (green) is calculated from a velocity range of
-52.5 to -48.5 km s$^{-1}$
with levels of
2.5 $\times$
0.076 K km s$^{-1}$(1$\sigma$).
}
\end{figure*}

\begin{figure*}
\includegraphics[angle=0,scale=0.6]{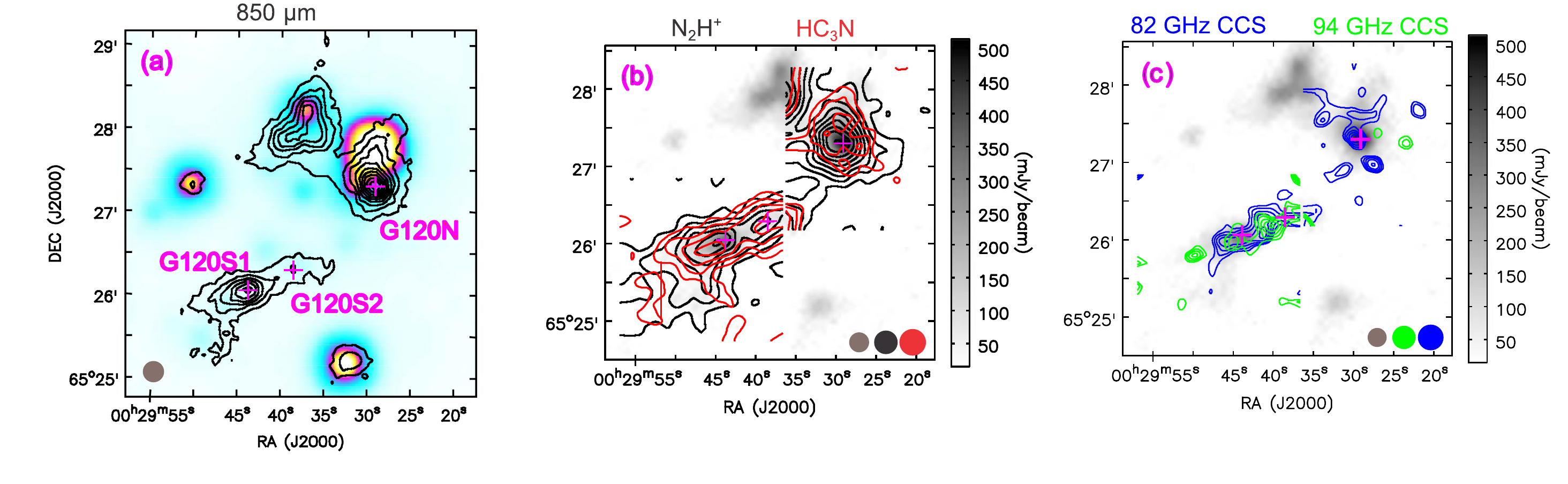}
\caption{Same as Figure 1 but for 
G120.7+2.7.
(a) The 850 $\mu$m continuum (contours) levels are
[0.1, 0.2, 0.3, 0.4, 0.5, 0.6, 0.7, 0.8, 0.9] $\times$
514.4 mJy/beam.
(b) 
For G120N,
The N$_2$H$^+$ intensity (black contours) is calculated from a velocity range of
-19.5 to -15.5 km s$^{-1}$
with levels of
[3, 6, 9, 12, 15, 18, 21, 24, 27] $\times$
0.098 K km s$^{-1}$(1$\sigma$).
For G120S,
The N$_2$H$^+$ intensity (black contours) is calculated from a velocity range of
-19.5 to -16.5 km s$^{-1}$
with levels of
[2.5, 3, 3.5, 4, 4.5, 5, 6, 7] $\times$
0.099 K km s$^{-1}$(1$\sigma$).
For G120N,
The HC$_3$N intensity (red) is calculated from a velocity range of
-19.5 to -15.5 km s$^{-1}$
with levels of
[3, 4.5, 6] $\times$
0.090 K km s$^{-1}$(1$\sigma$).
For G120S,
The HC$_3$N intensity (red) is calculated from a velocity range of
-19.5 to -16.5 km s$^{-1}$
with levels of
[3, 6, 9, 12] $\times$
0.062 K km s$^{-1}$(1$\sigma$).
(c)
For G120N,
The 82 GHz CCS intensity (blue) is calculated from a velocity range of
-19.5 to -15.5 km s$^{-1}$
with levels of
[2.5, 3, 3.5, 4, 4.5] $\times$
0.1 K km s$^{-1}$(1$\sigma$).
For G120S,
The 82 GHz CCS intensity (blue) is calculated from a velocity range of
-19.5 to -16.5 km s$^{-1}$
with levels of
[3, 3.5, 4, 4.5, 5] $\times$
0.080 K km s$^{-1}$(1$\sigma$).
For G120N,
The 94 GHz CCS intensity (green) is calculated from a velocity range of
-19.5 to -15.5 km s$^{-1}$
with levels of
[2.5, 3, 3.5] $\times$
0.11 K km s$^{-1}$(1$\sigma$).
For G120S,
The 94 GHz CCS intensity (green) is calculated from a velocity range of
-19.5 to -16.5 km s$^{-1}$
with levels of
[3, 3.5, 4, 4.5, 5] $\times$
0.085 K km s$^{-1}$(1$\sigma$).
}
\end{figure*}

\begin{figure*}
\includegraphics[angle=0,scale=0.4]{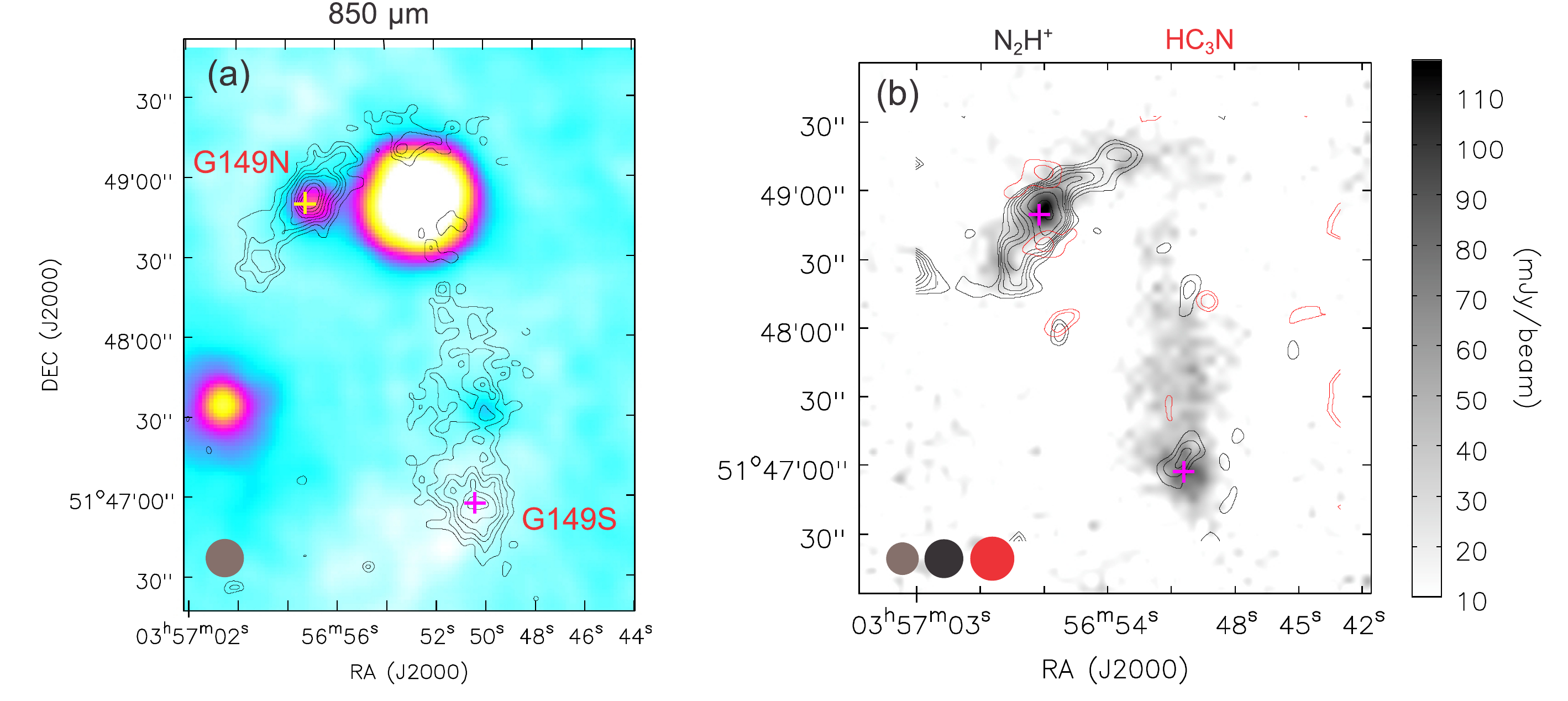}
\caption{Same as Figure 1 but for G149.5-1.2.
(a) The 850 $\mu$m continuum (contours) levels are
[0.2, 0.3, 0.4, 0.5, 0.6, 0.7, 0.8, 0.9] $\times$
117.2 mJy/beam.
(b) The N$_2$H$^+$ intensity (black contours) is calculated from a velocity range of
-8.5 to -5.5 km s$^{-1}$
with levels of
[2.5, 3, 3.5, 4, 4.5, 5, 6, 7] $\times$
0.099 K km s$^{-1}$(1$\sigma$).
The HC$_3$N intensity (red) is calculated for  a velocity range of
-8.5 to -5.5 km s$^{-1}$
with levels of
[2.5, 3] $\times$
0.080 K km s$^{-1}$(1$\sigma$).
}
\end{figure*}

\begin{figure*}
\includegraphics[angle=0,scale=0.4]{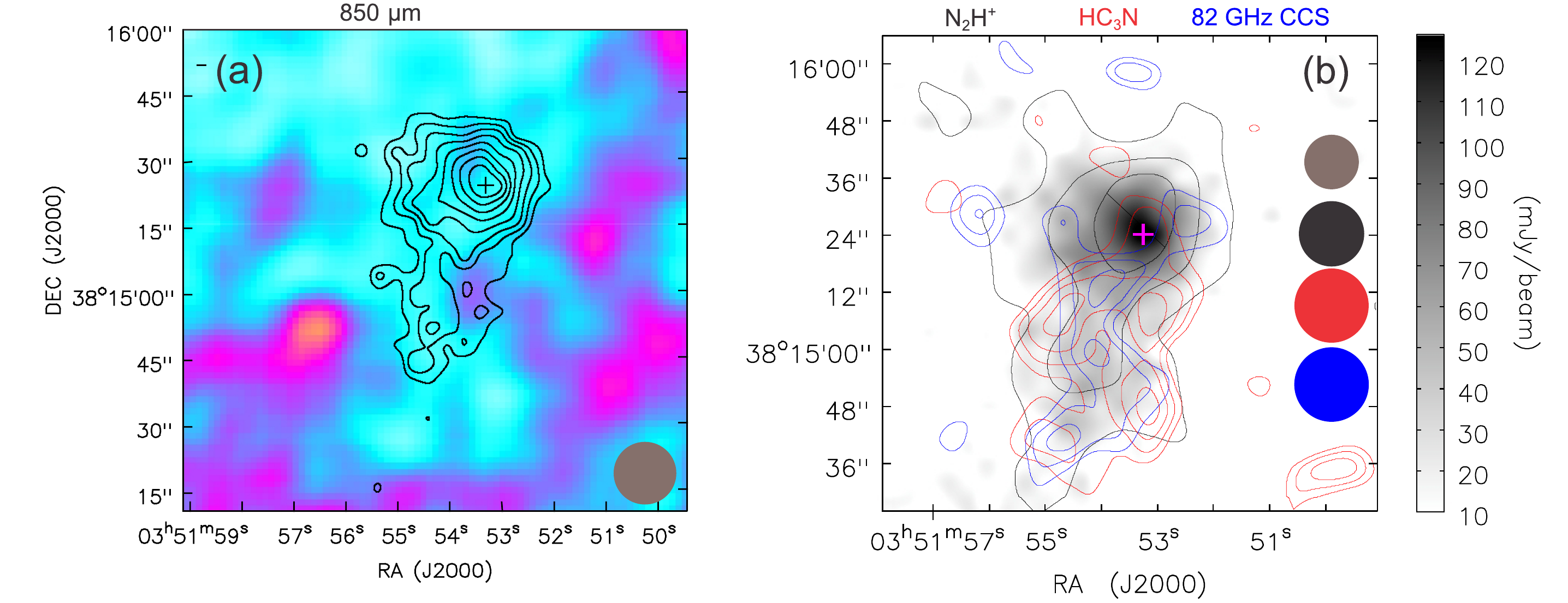}
\caption{Same as Figure 1 but for G157.6-12.2.
(a) The 850 $\mu$m continuum (contours) levels are
[0.2, 0.3, 0.4, 0.5, 0.6, 0.7, 0.8, 0.9] $\times$
126.8 mJy/beam.
(b)
The N$_2$H$^+$ intensity (black contours) is calculated from a velocity range of
-8.5 to -6.5 km s$^{-1}$
with levels of
[3, 6, 9] $\times$
0.078 K km s$^{-1}$(1$\sigma$).
The HC$_3$N intensity (red) is calculated from a velocity range of
-8.5 to -6.5 km s$^{-1}$
with levels of
[2.5, 3, 3.5, 4] $\times$
0.071 K km s$^{-1}$(1$\sigma$).
The 82 GHz CCS intensity (blue) is calculated from a velocity range of
-8.5 to -6.5 km s$^{-1}$
with levels of
[2.5, 3, 3.5] $\times$
0.086 K km s$^{-1}$(1$\sigma$).
}
\end{figure*}

\begin{figure*}
\includegraphics[angle=0,scale=0.4]{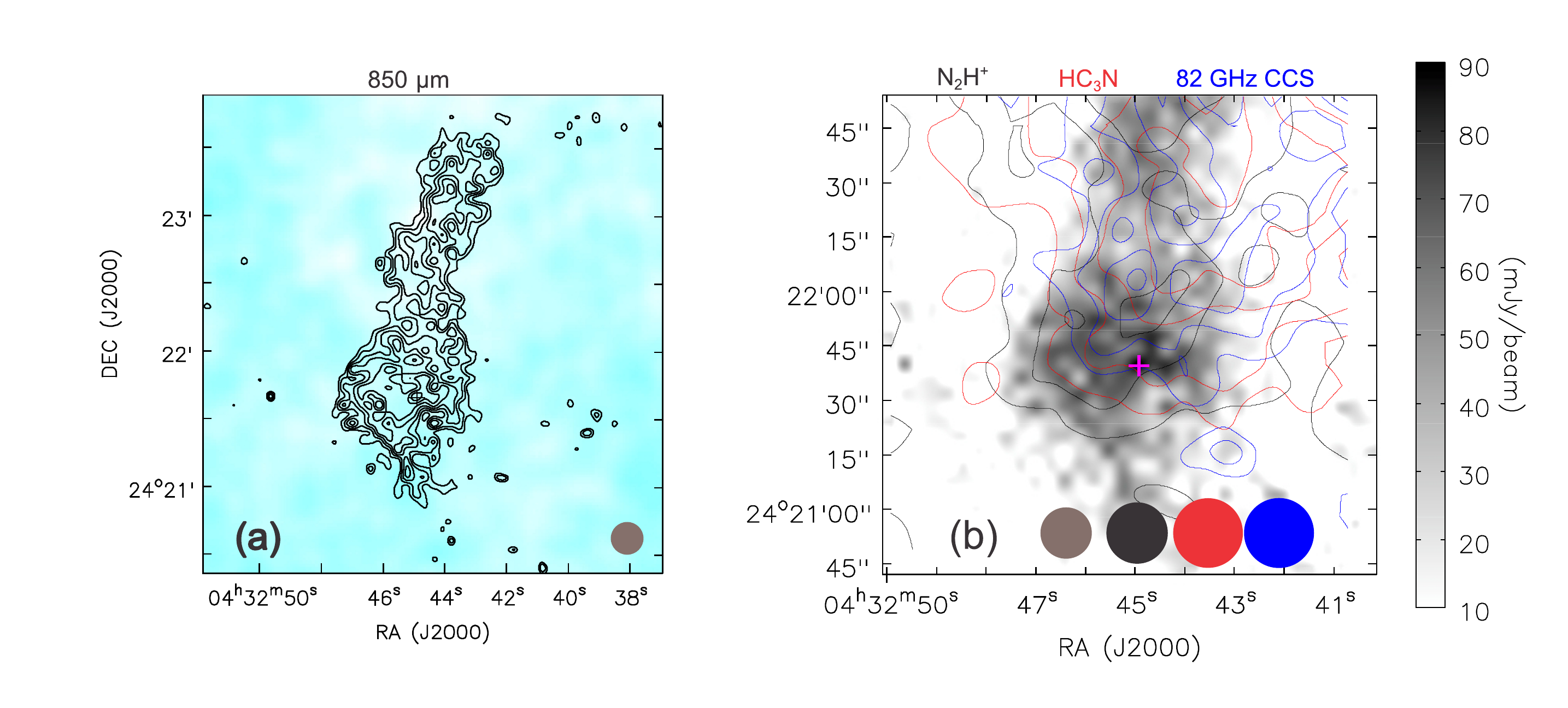}
\caption{Same as Figure 1 but for G174.0-15.8.
(a) The 850 $\mu$m continuum (contours) levels are
[0.3, 0.4, 0.5, 0.6, 0.7, 0.8, 0.9] $\times$
90.3 mJy/beam.
(b) 
Black contours representintensity integrated for a velocity range of
5.5 to 7.5 km s$^{-1}$
with levels of
[3, 4.5, 6] $\times$
0.065 K km s$^{-1}$(1$\sigma$).
The HC$_3$N intensity (red) is calculated from a velocity range of
5.5 to 7.5 km s$^{-1}$
with levels of
[3, 6, 9, 12] $\times$
0.052 K km s$^{-1}$(1$\sigma$).
The 82 GHz CCS intensity (blue) is calculated from a velocity range of
5.5 to 7.5 km s$^{-1}$
with levels of
[3, 4.5, 6] $\times$
0.065 K km s$^{-1}$(1$\sigma$).
}
\end{figure*}

\begin{figure*}
\includegraphics[angle=0,scale=0.4]{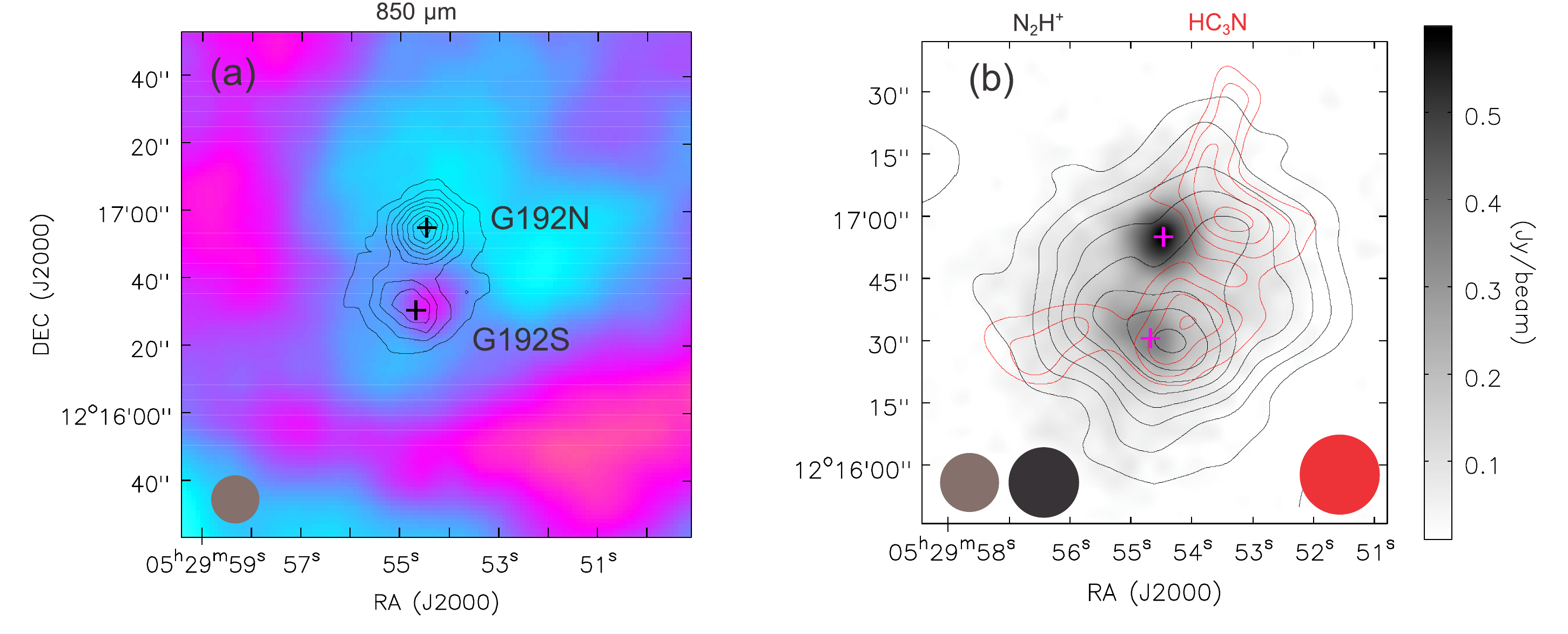}
\caption{Same as Figure 1 but for G192.32-11.88.
(a) The 850 $\mu$m continuum (contours) levels are
[0.2, 0.3, 0.4, 0.5, 0.6, 0.7, 0.8, 0.9] $\times$
600 mJy/beam.
(b) 
The N$_2$H$^+$ intensity (black contours) is calculated from a velocity range of
11.5 to 13.5 km s$^{-1}$
with levels of
[3, 6, 9, 12, 15, 18, 21, 24] $\times$
0.081 K km s$^{-1}$(1$\sigma$).
The HC$_3$N intensity (red) is calculated from a velocity range of
11.5 to 13.5 km s$^{-1}$
with levels of
[3, 4, 5, 6] $\times$
0.069 K km s$^{-1}$(1$\sigma$).
}
\end{figure*}

\begin{figure}
\includegraphics[angle=0,scale=0.3]{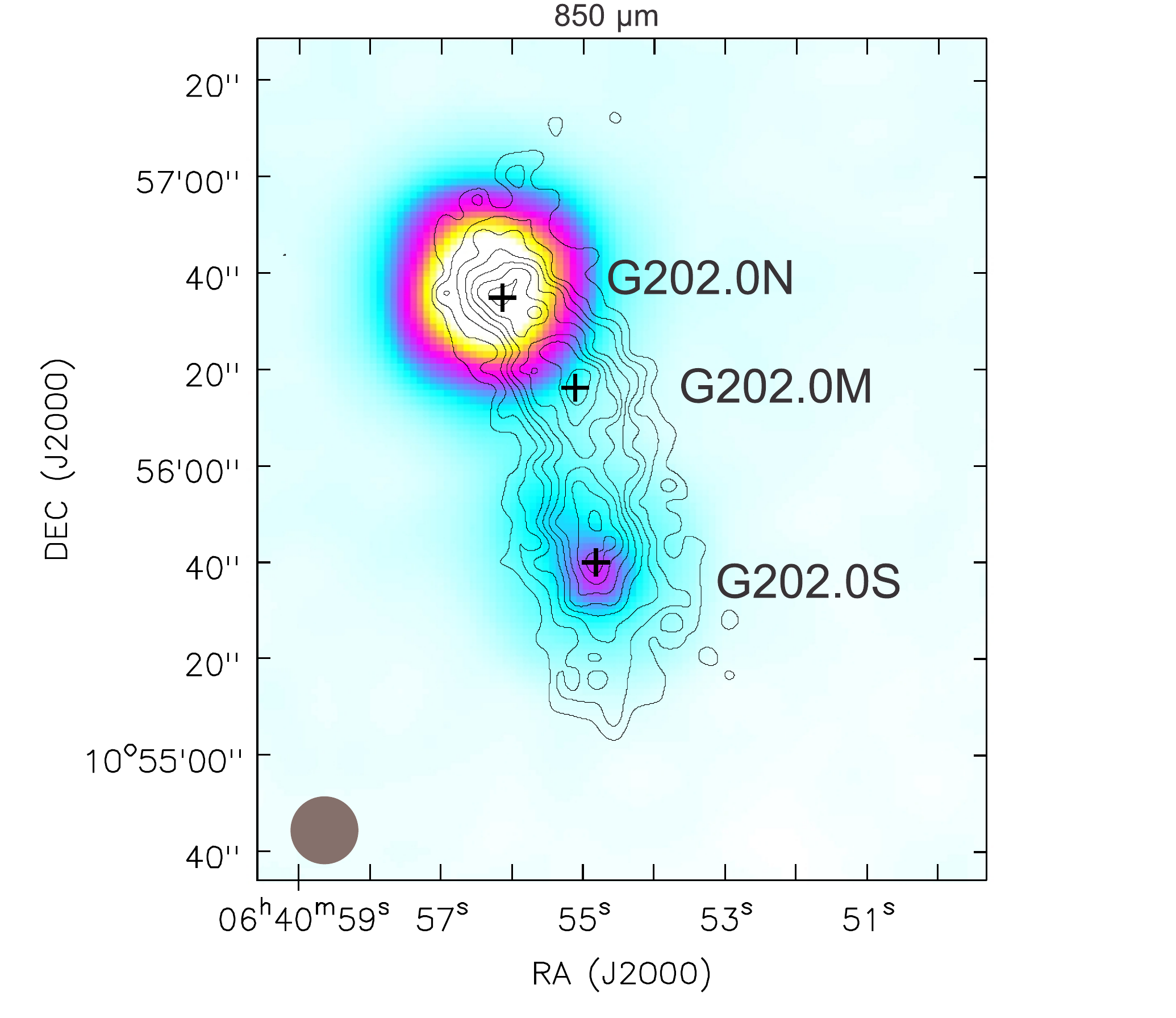}
\caption{The 850 $\mu$m continuum image (contours) obtained with JCMT/SCUBA-2 toward 
G202.00+2.65
superimposed on the WISE 22 $\mu$m image (color).
Contour levels are
[0.2, 0.3, 0.4, 0.5, 0.6, 0.7, 0.8, 0.9] $\times$
88.8 mJy/beam.
}
\end{figure}

\begin{figure}
\includegraphics[angle=0,scale=0.3]{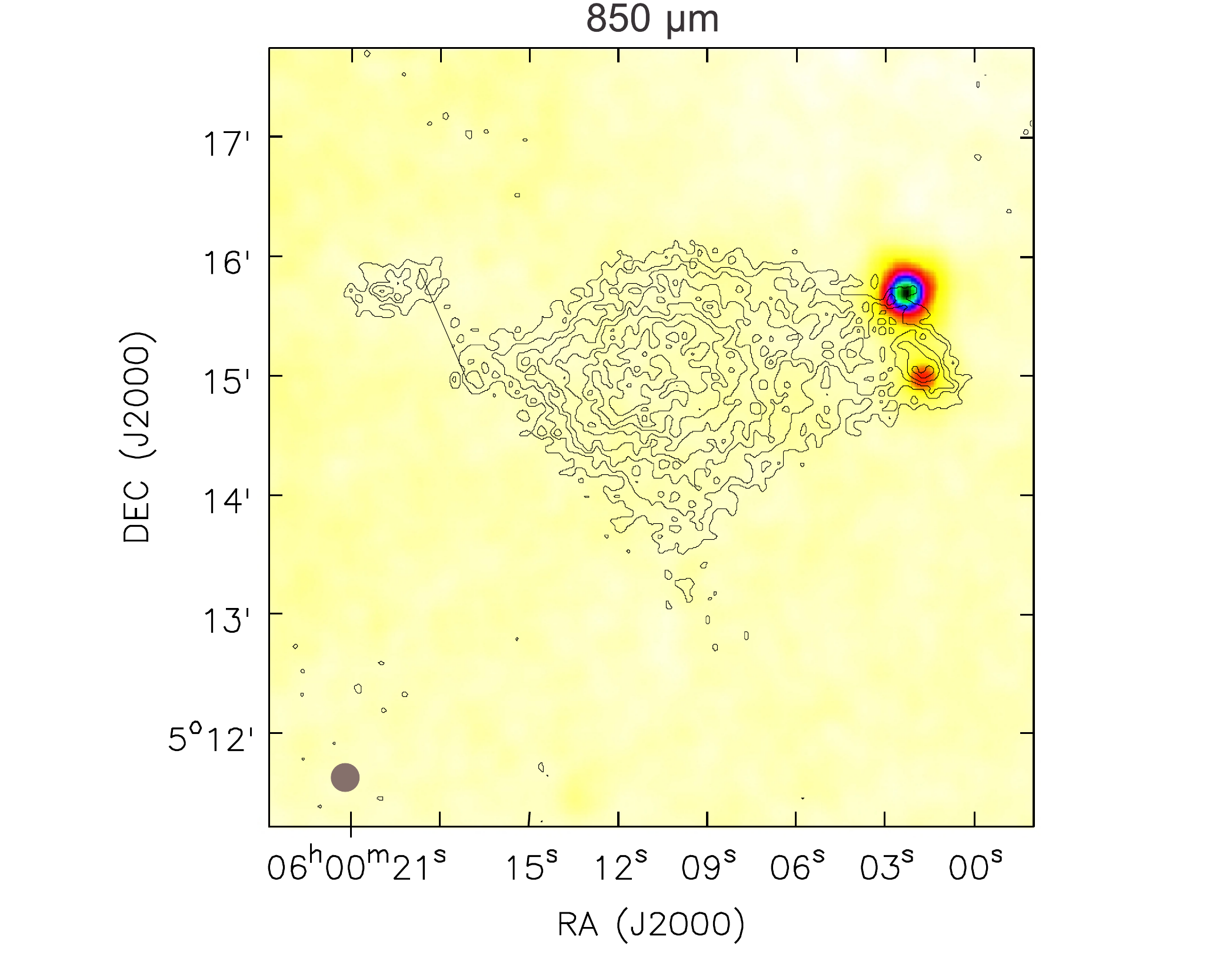}
\caption{Same as Figure 8 but for G202.31-8.92.
Contour levels are
[0.2, 0.3, 0.4, 0.5, 0.6, 0.7, 0.8, 0.9] $\times$
231.2 mJy/beam.
}
\end{figure}

\begin{figure*}
\includegraphics[angle=0,scale=0.4]{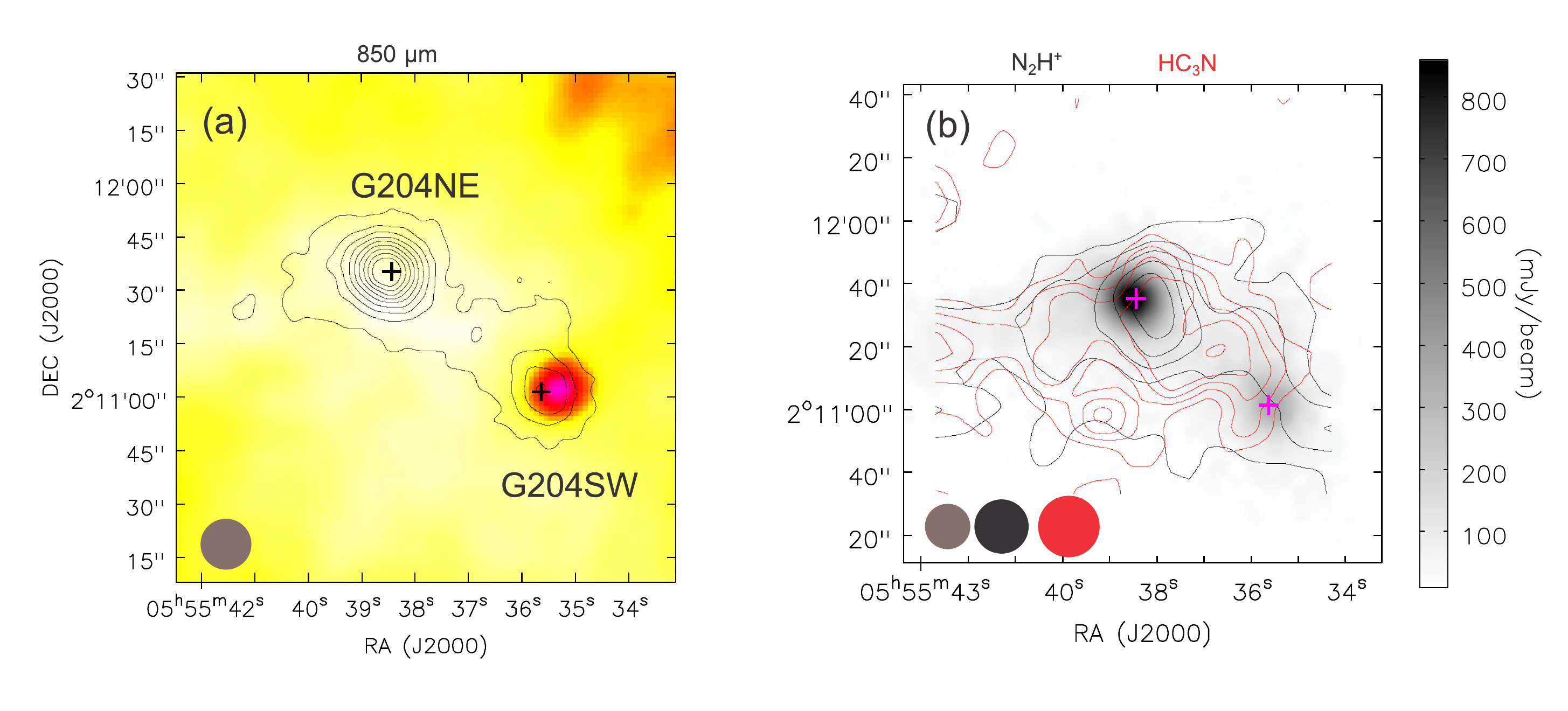}
\caption{Same as Figure 1 but for G204.4-11.3.
(a) The 850 $\mu$m continuum (contours) levels are
[0.1, 0.2, 0.3, 0.4, 0.5, 0.6, 0.7, 0.8, 0.9] $\times$
861.5 mJy/beam.
(b) The N$_2$H$^+$ intensity (black contours) is calculated from a velocity range of
0.5 to 2.5 km s$^{-1}$
with levels of
[3, 6, 9, 12, 15] $\times$
0.11 K km s$^{-1}$(1$\sigma$).
The HC$_3$N intensity (red) is calculated for 
intensity integrated for a velocity range of
0.5 to 2.5 km s$^{-1}$
with levels of
[3, 4.5, 6, 7.5, 10] $\times$
0.086 K km s$^{-1}$(1$\sigma$).
}
\end{figure*}

\begin{figure*}
\includegraphics[angle=0,scale=0.4]{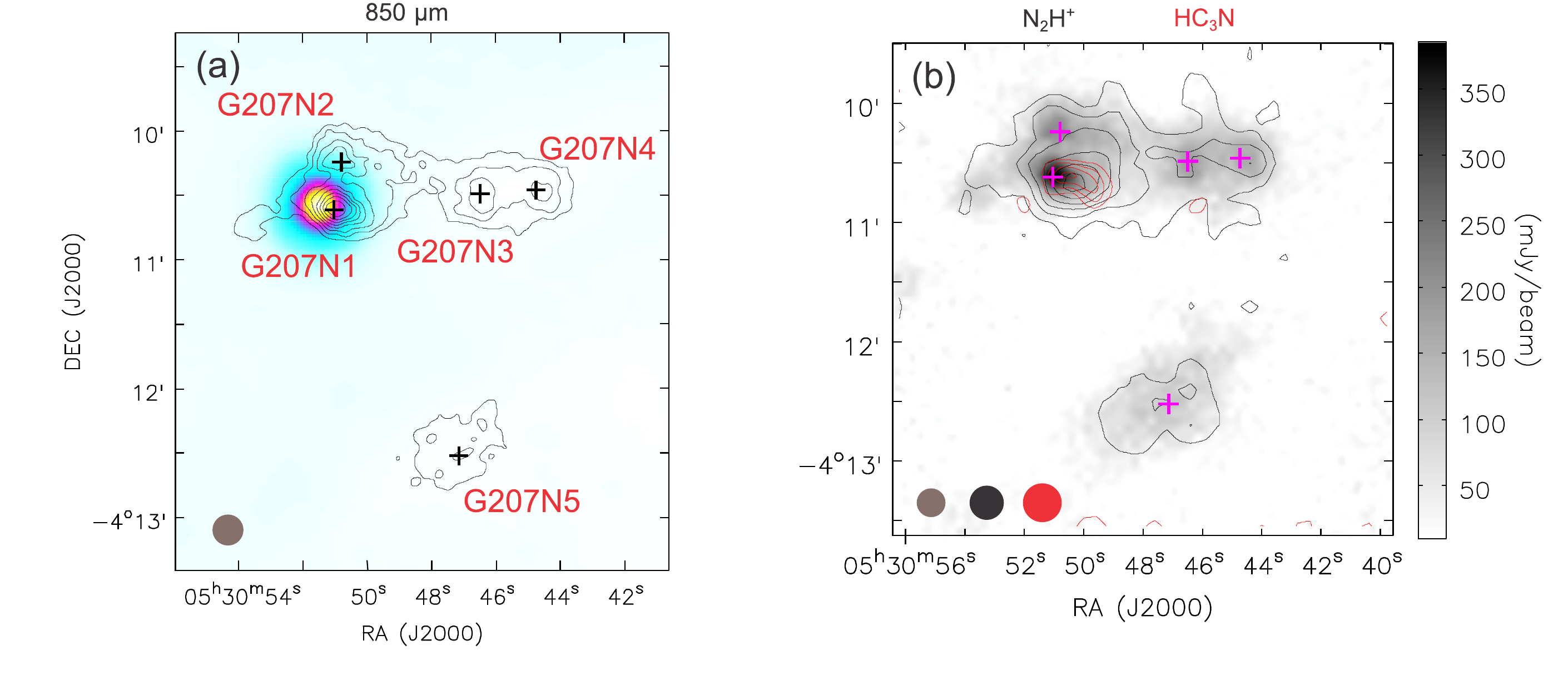}
\caption{Same as Figure 1 but for G207N.
(a) The 850 $\mu$m continuum (contours) levels are
[0.1, 0.2, 0.3, 0.4, 0.5, 0.6, 0.7, 0.8, 0.9] $\times$
386.4 mJy/beam.
(b) 
The N$_2$H$^+$ intensity (black contours) is calculated from a velocity range of
9.5 to 12.5 km s$^{-1}$
with levels of
[3, 6, 9, 12, 15, 18] $\times$
0.13 K km s$^{-1}$(1$\sigma$).
The HC$_3$N intensity (red) is calculated for
intensity integrated for a velocity range of
9.5 to 12.5 km s$^{-1}$
with levels of
[3, 4.5, 6, 7.5] $\times$
0.11 K km s$^{-1}$(1$\sigma$).
}
\end{figure*}

\begin{figure*}
\includegraphics[angle=0,scale=0.4]{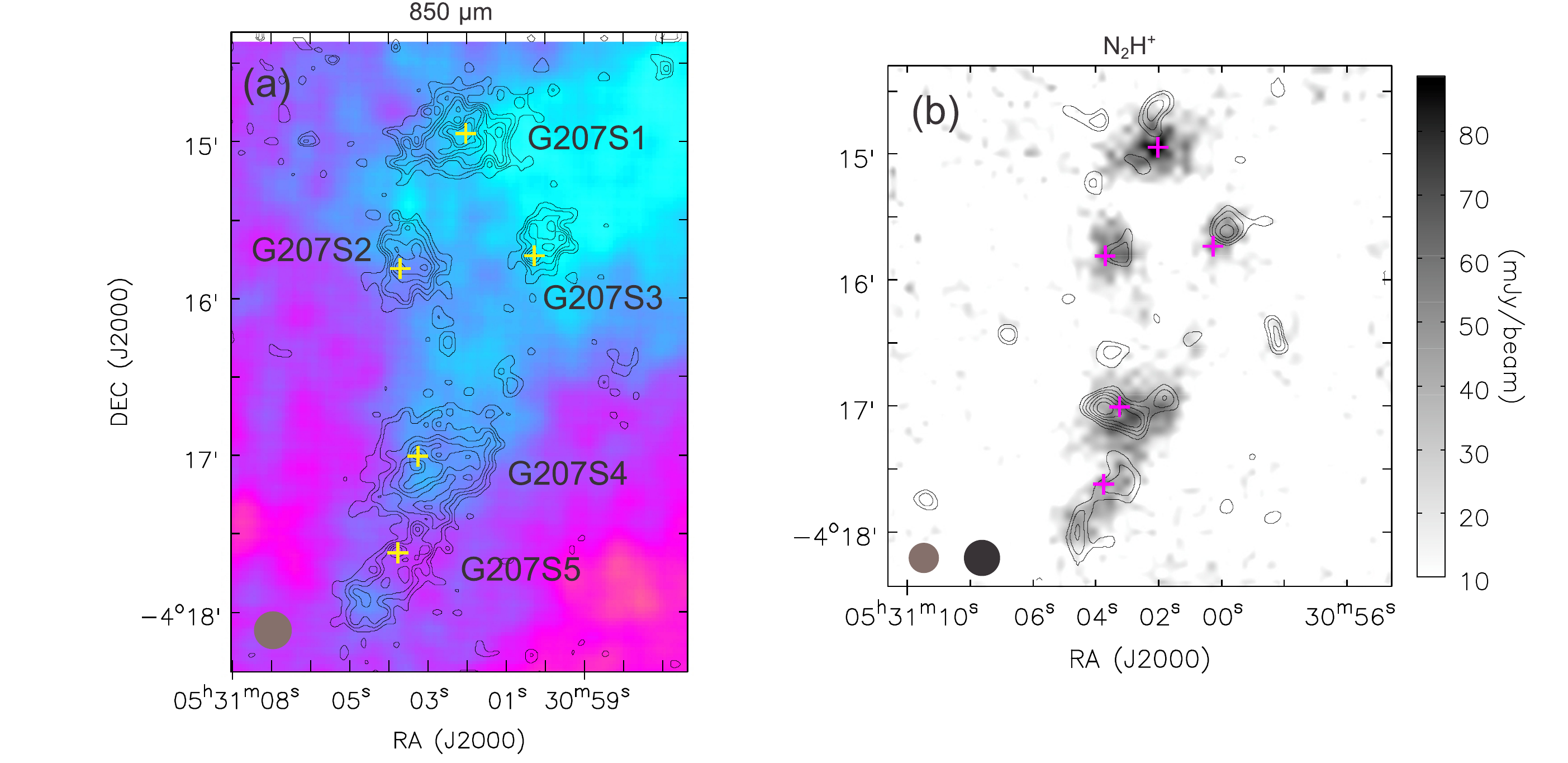}
\caption{Same as Figure 1 but for G207S.
(a)The 850 $\mu$m continuum (contours) levels are
[0.2, 0.3, 0.4, 0.5, 0.6, 0.7, 0.8, 0.9] $\times$
88.8 mJy/beam.
(b) The N$_2$H$^+$ intensity (black contours) is calculated from a velocity range of
11.5 to 12.5 km s$^{-1}$
with levels of
[2.5, 3, 3.5, 4, 4.5, 5] $\times$
0.083 K km s$^{-1}$(1$\sigma$).
}
\end{figure*}

\begin{figure*}
\includegraphics[angle=0,scale=0.6]{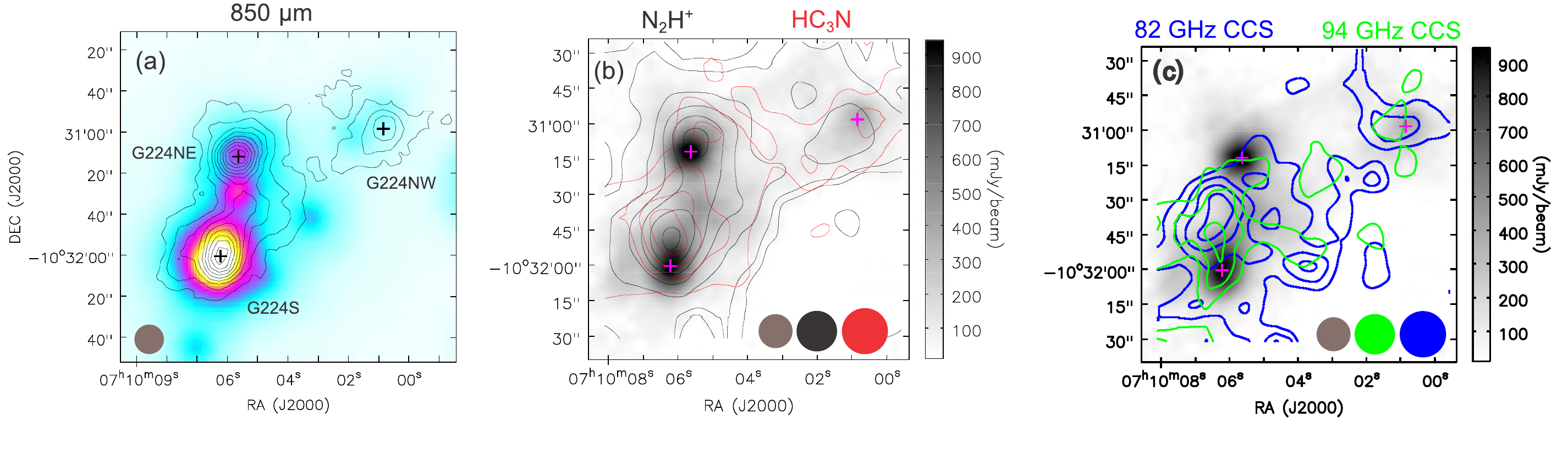}
\caption{Same as Figure 1 but for G224.4-0.6.
(a) The 850 $\mu$m continuum (contours) levels are
[0.1, 0.2, 0.3, 0.4, 0.5, 0.6, 0.7, 0.8, 0.9] $\times$
946.5 mJy/beam.
(b)
The N$_2$H$^+$ intensity (black contours) is calculated from a velocity range of
12.5 to 17.5 km s$^{-1}$
with levels of
[3, 5, 10, 15, 20, 25, 30] $\times$
0.15 K km s$^{-1}$(1$\sigma$).
The HC$_3$N intensity (red) is calculated from a velocity range of
12.5 to 17.5 km s$^{-1}$
with levels of
[3, 6, 9, 12] $\times$
0.12 K km s$^{-1}$(1$\sigma$).
 (c) The 82 GHz CCS intensity (blue) is calculated for  a velocity range of
12.5 to 17.5 km s$^{-1}$
with levels of
[3, 4.5, 6, 7.5] $\times$
0.13 K km s$^{-1}$.
The 94 GHz CCS intensity (green) is calculated from a velocity range of
12.5 to 17.5 km s$^{-1}$
with levels of
[3, 4.5, 6] $\times$
0.14 K km s$^{-1}$(1$\sigma$).
}
\end{figure*}

\section{Individual Objects}

\subsection{G089.9-01.9}

The object is located in L974 \citep{1962ApJS....7....1L,2005PASJ...57S...1D}, and in the dark cloud Khavtassi 137 \citep{kha60}.  
There is a Class 0 like source (IRAS 21182+4611), which is a bright source in the WISE image, northeast of the western core G089W with an offset of $\sim$ 15$\arcsec$.  
Then, we regard G089W itself as starless.  
To our knowledge, there is no information that the eastern core G089E is associated with any young stellar object, and
we also regard it as starless.
Figure 1 shows maps toward G089.9-01.9.
The N$_2$H$^+$ distribution shows a resemblance to the 850 $\mu$m distribution, but their peak positions are slightly different.
The HC$_3$N distribution shows some similarity to the 850 $\mu$m distribution, but the correlation between their intensities is poor.
The 82 GHz CCS distribution is very clumpy, and appears as if it surrounds the 850 $\mu$m core.
The 94 GHz CCS distribution is different, but it is still located on both sides of the 850 $\mu$m core.
Nobeyama T70 observations were made toward G089W.  The column density ratio of $N$(DNC) to $N$(HN$^{13}$C) is 4.5, which is close to the value of 3.0 at L1544 \citep{2003ApJ...594..859H,2006ApJ...646..258H}.
L1544 is known as a prestellar core showing collapsing motion \citep{1998ApJ...504..900T}.

\subsection{G108.8-00.8}

G108.8-00.8 is located in a GMC (1$\arcdeg \times 15\arcmin$) associated with five \citet{1959ApJS....4..257S} H\textsc{ii} regions, S147, S148, S149, S152, and S153, and also associated with the supernova remnant G109.1-1.0 (CTB109) in the Perseus arm \citep{1985PASJ...37..345T,1987AA...184..279T}.
These two cores are located between peaks $\alpha$ (corresponding to S152) and $\zeta$ in \citet{1985PASJ...37..345T}.
G108.8-00.8S is associated with two Class I-like sources seen in the WISE image, while G108.8-00.8N is starless.
Figure 2 shows maps toward G108.8-00.8.
The N$_2$H$^+$ distribution resembles the 850 $\mu$m distribution.
The HC$_3$N emission is observed at the edge of the 850 $\mu$m core of G108N
, and the correlation between N$_2$H$^+$ and HC$_3$N is poor.  For G108S, their distributions are more or less correlated.  The 82 GHz CCS emission is detected toward the N$_2$H$^+$/850 $\mu$m cores (G108N and G108S) and also the edge of them (G108N).
Note that this source is distant, the spatial resolution is as large as 0.3 pc.  
It is possible that different distributions between N$_2$H$^+$ and CCS are less clear due to its large distance.
The kinetic temperature of G108S is 14.3$\pm$3.0 K.
G108S is as cold as typical cold dark clouds.
Nobeyama T70 observations were made toward G108N, a starless core.  
We detected DNC, but not HN$^{13}$C. The column density ratio of $N$(DNC) to $N$(HN$^{13}$C) is $>$ 2.5, which is similar to the value of 3.0 at L1544 \citep{2003ApJ...594..859H,2006ApJ...646..258H}.
See also \citet{kim2016} for our observations with other telescopes such as PMO 14m, CSO, IRAM 30 m telescopes, etc.

\subsection{G120.7+2.7}


Figure 3 shows maps toward G120.7+2.7.
The WISE source is clearly  offset from G120N, and we regard the latter as a starless core.
Both  N$_2$H$^+$ and HC$_3$N distributions are well correlated with the 850 $\mu$m distribution.
In G120N, the 82 GHz CCS emission has two intensity peaks, and one of them is close to the N$_2$H$^+$ emission peak. Toward the N$_2$H$^+$ peak, we detected N$_2$D$^+$.
In G120S, the N$_2$H$^+$ emission shows two cores.
We clearly detected both 82GHz and 94GHz CCS emission at G120S.
It is possible that the CCS emission surrounds the N$_2$H$^+$ cores, although it is less clear.
Nobeyama T70 observations were made toward G120N, a starless core.  The column density ratio of $N$(DNC) to $N$(HN$^{13}$C) is 1.7, which is close to the value of 1.91 at L1498 \citep{2006ApJ...646..258H}.

\subsection{G149.5-1.2}

This clump is located at the edge of the dark cloud Khavtassi 241 \citep{kha60}.
Figure 4 shows maps toward G149.5-1.2.
The N$_2$H$^+$ distribution is well correlated with the 850 $\mu$m distribution.
The HC$_3$N emission is poorly correlated with the 850 $\mu$m distribution, and it is located on both sides of the north-eastern 850 $\mu$m (and N$_2$H$^+$) core.
We detected the N$_2$D$^+$ emission in single pointing observations
between the brightest N$_2$H$^+$ core G149N (WISE source) and the northern HC$_3$N core.
The column density ratio of $N$(DNC) to $N$(HN$^{13}$C) is 3.2, which is close to the value of 3.0 in L1544
\citep{2003ApJ...594..859H,2006ApJ...646..258H}.

\subsection{G157.6-12.2}

This clump is located near the dark cloud 
L1449 \citep{1962ApJS....7....1L,2005PASJ...57S...1D}
and in 
Khavtassi 257 \citep{kha60}, which is close to the California Nebula, NGC 1499.
Figure 5 shows maps toward G157.6-12.2.
The N$_2$H$^+$ distribution is well correlated with the 850 $\mu$m distribution.
The 82 GHz CCS emission surrounds the 850 $\mu$m (and N$_2$H$^+$) core.
The HC$_3$N emission shows distribution different from the N$_2$H$^+$ distribution, and  looks anticorrelated with the 82 GHz CCS emission.
Nobeyama T70 observations were made toward the center of G157, a starless core.  The column density ratio of $N$(DNC) to $N$(HN$^{13}$C) is 6.9, which is larger than the value of 3.0 at L1544
\citep{2003ApJ...594..859H,2006ApJ...646..258H}.

\subsection{G174.0-15.8}

This clump is located in 
L1529 \citep{1962ApJS....7....1L,1985AApS...60...43W,2005PASJ...57S...1D} in Taurus.
G174.0-15.8 is a starless clump.
Figure 6 shows maps toward  G174.0-15.8.
The N$_2$H$^+$ distribution is more or less correlated with the 850 $\mu$m distribution.
The HC$_3$N and 82 GHz CCS emission is distributed more extensively than the 850 $\mu$m core.

\subsection{G192.32-11.88}

G192.3-11.8 is located in the $\lambda$ Orionis complex, and is associated with B30 cataloged by \citet{1927pasr.book.....B}, 
L1581 \citep{1962ApJS....7....1L,2005PASJ...57S...1D}, and 
with No. 9 CO emission peak identified by \citet{1986ApJ...303..375M}. It is located in the dark cloud Khavtassi 296 \citep{kha60}.
\citet{2016ApJS..222....7L} discovered
an extremely young Class 0 protostellar object (G192N) and a proto-brown dwarf candidate (G192S).
Observations with SMA show the existence of molecular outflows associated with these objects.
Figure 7 shows maps toward G192.32-11.88.
The N$_2$H$^+$ distribution shows an intensity peak toward G192S, while there is no N$_2$H$^+$
peak corresponding to G192N. 
The 850 $\mu$m map of SCUBA-2 has two clear peaks, G192N being the more intense one.
The HC$_3$N distribution is very different from that of 850 $\micron$ (and N$_2$H$^+$), and is clumpy.
G192S is associated with one HC$_3$N core, while there is no HC$_3$N counterpart to G192N.
It seems that the 850 $\micron$ (and N$_2$H$^+$) peak is surrounded by HC$_3$N cores.

\citet{2016ApJS..222....7L} detected large velocity gradient in this region in $^{13}$CO $J$ = 1$-$0 and 2$-$1 in the NE-SW direction, and attributed its origin to compression by the H$\textsc{ii}$ region. 
The N$_2$H$^+$ core inside shows an E-W velocity gradient of the order of 0.5 km s$^{-1}$ arcmin $^{-1}$ or 4 km s$^{-1}$ pc $^{-1}$. This gradient is consistent with what was observed by \citet{2016ApJS..222....7L} toward the center of their Clump-S in $^{13}$CO $J$ = 1$-$0 and 2$-$1 (see their Figure 2).
Figure 14 shows the intensity-weighted radial velocity (moment 1) map toward G192.32-11.88 in the N$_2$H$^+$
main hyperfine emission
group $F_1$ = 2$-$1 ($F$ =  1$-$0,  2$-$1, and 3$-$2).
This shows a velocity gradient of 0.5 km s$^{-1}$ across the clump in the E-W direction.
Because this object is located in the Orion region, where the specific angular momentum was statistically investigated by \citet{2016PASJ...68...24T},
and also because this core shows a clear velocity gradient,
we investigate its properties in detail. 
It is well known that  the specific angular momentum  (angular momentum per unit mass) $J/M$ is roughly proportional to $R^{1.6}$ for molecular clouds and their cores having sizes of 0.1 to 30 pc
in general \citep{1985prpl.conf..137G,1993ApJ...406..528G,1995ARAA..33..199B}.
We compare the radius, velocity gradient, and the specific angular momentum
among these emission lines.
The beam-corrected half-intensity radius $R$ is 0.26, 0.16, and 0.058 pc, the velocity gradient is 3.9, 2.2, and 0.23 km s$^{-1}$, and $J/M$ is $1.3\times 10^{23}$, $4.5\times 10^{22}$, and $1.7\times 10^{21}$ cm$^2$ s$^{-1}$, respectively, in $^{13}$CO $J$ = 1$-$0 and 2$-$1 \citep{2016ApJS..222....7L} and N$_2$H$^+$ (this study).
\citet{2016PASJ...68...24T} investigated the specific angular momentum $J/M$ of N$_2$H$^+$ cores in the Orion A GMC, and compared them with cold cloud cores observed in NH$_3$ by \citet{1993ApJ...406..528G}.  
Figure 15 plots the specific angular momentum $J/M$ of G192.32-11.88 cores observed in this study as well as \citet{2016PASJ...68...24T,1993ApJ...406..528G}..  $J/M$ of G192.32-11.88 in N$_2$H$^+$ is within the range found for cores in the Orion A GMC, but is located at the high end of it.  It is possible that compression by the H$\textsc{ii}$ region has resulted in relatively large $J/M$ found in the present study.    

Nobeyama T70 observations were made toward G192N associated with the Class 0 like protostar.  The column density ratio of $N$(DNC) to $N$(HN$^{13}$C) is 5.8, which is larger than the value of 3.0 at L1544
\citep{2003ApJ...594..859H,2006ApJ...646..258H}.
Detection of DNC and N$_2$D$^+$ indicates that this core is chemically relatively young, although the core is star forming.

\begin{figure}
\includegraphics[angle=0,scale=0.3]{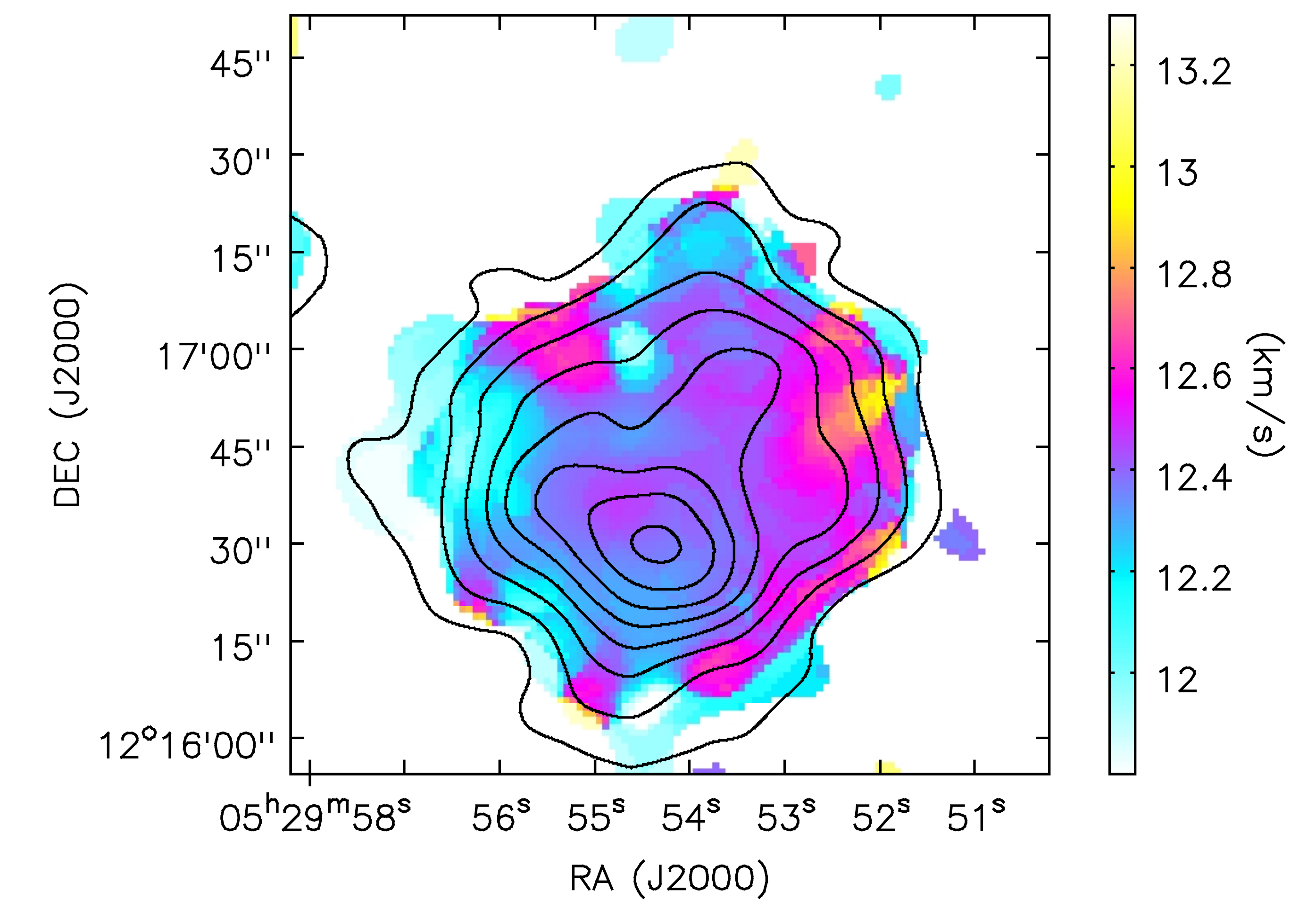}
\caption{The intensity-weighted radial velocity  (moment 1) map (color) toward G192.32-11.88 in the N$_2$H$^+$ $J = 1-0$
$F_1$ = 2$-$1 main hyperfine emission group  $F_1$ = 2$-$1 ($F$ =  1$-$0,  2$-$1, and 3$-$2)
overlaid on the integrated intensity (moment 0) contour map.
The emission less than 2$\sigma$ is ignored in the moment 1 calculation.
The contour levels are the same as those in Figure 7.}
\end{figure}

\begin{figure}
\includegraphics[angle=0,scale=0.5]{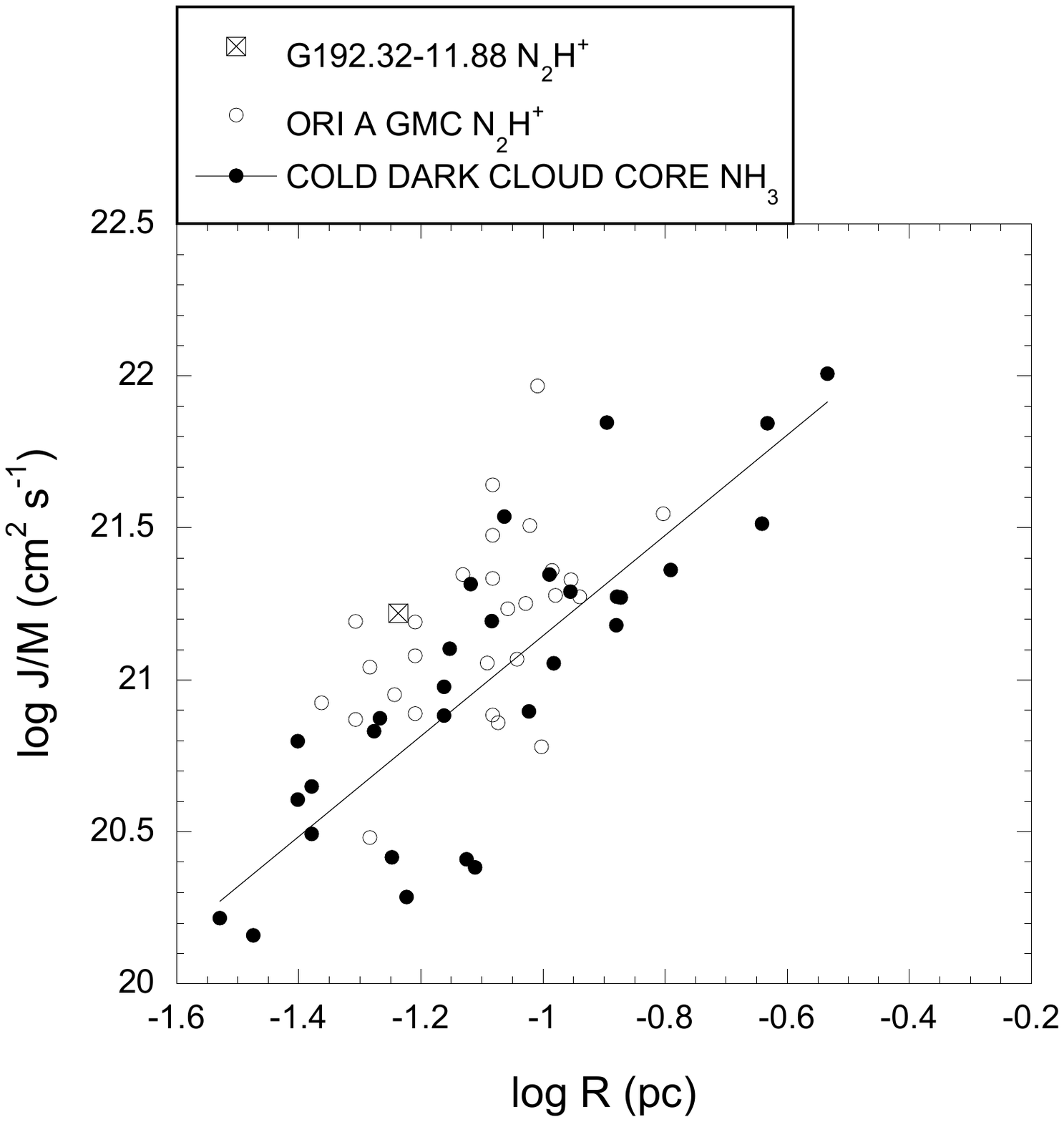}
\caption{$J/M-R$ diagram for G192.32-11.88 in N$_2$H$^+$ in this study (box with cross), Orion A GMC cores (open circles, \citet{2016PASJ...68...24T}), and cold dark
cloud cores (filled circles, \citet{1993ApJ...406..528G}).
The thin solid straight line
is computed using a linear least-squares program for 
cold dark cloud cores. }
\end{figure}

\subsection{G202.00+2.65}

This clump is located in the dark cloud 
L1613 \citep{1962ApJS....7....1L,2005PASJ...57S...1D} 
and Khavtassi 308 \citep{kha60}.
$T_k$ is derived to be $<$10.2 K.
Figure 8 shows the continuum map.
The upper limit to the kinetic temperature (10.2 K) was obtained from NH$_3$ observations.

\subsection{G202.31-8.92}

G202.31-8.92 is located in 
L1611 \citet{1962ApJS....7....1L,2005PASJ...57S...1D}
in the Orion Northern Filament, and close to No. 52 CO Emission Peak cataloged by \citet{1986ApJ...303..375M},
which has two CO velocity components 8.8 and 11.4 km s$^{-1}$.
As \citet{1986ApJ...303..375M} discussed, we assume that the distance to G202.31-8.92 is similar to that of Orion A and B GMCs.
Figure 9 shows the continuum map.
$T_k$ is derived to be $<$11.6 K.
Two velocity components (9.2 and 12.0 km~s$^{-1}$) were detected  in the $^{12}$CO (1-0) and  $^{13}$CO (1-0) emission toward this source \citep{2012ApJS..202....4L}. 
The continuum emission shows a morphology similar to that of the 12.0 km~s$^{-1}$ CO clump. 
The velocity of the NH$_3$ emission is 11.92 km~s$^{-1}$, which is similar to that of the 12.0 km~s$^{-1}$ CO clump. 

The upper limit to the kinetic temperature  (11.6 K) was obtained from NH$_3$ observations.

\subsection{G204.4-11.3}

G204.4-11.3 is located in 
L1621 \citet{1962ApJS....7....1L,2005PASJ...57S...1D} 
and the Orion B GMC, and close to No.37 CO Emission Peak cataloged by 
\citet{1962ApJS....7....1L,1986ApJ...303..375M}, 
which has two CO velocity components 8.6 and 10.9 km s$^{-1}$.
It is located near the edge of the dark cloud Khavtassi 311 \citep{kha60}.
As \citet{1986ApJ...303..375M} discussed, we assume that the distance to G204.4-11.3 is similar to that of Orion A and B GMCs.
Figure 10 shows maps toward G204-11.3.
The N$_2$H$^+$ emission shows a peak toward the starless peak G204NE in the 850 $\mu$m continuum, 
but with a marginal offset of $\sim10\arcsec$.
The HC$_3$N is more extended than 850 $\mu$m (and N$_2$H$^+$).
Toward G204NE (virtually identical to the T70 position), we detected the 82 GHz CCS emission (Tables 4 and 5), but the emission is too weak and narrow to draw a reliable map.
In the single pointing observations toward G204NE, we have detected both DNC and N$_2$D$^+$.
The column density ratio of $N$(DNC) to $N$(HN$^{13}$C) is 15.0, which is much larger than the value of 3.0 at L1544
\citep{2003ApJ...594..859H,2006ApJ...646..258H}.
The high deuterium fraction ratio implies that G204NE is a starless core on the verge of star formation.
Figure 16 shows the moment 1 radial velocity map toward G204-11.3 in N$_2$H$^+$.
We do not see a prominent velocity gradient ($<$ 0.5 km s$^{-1}$ arcmin$^{-1}$ or $<$4 km s$^{-1}$ pc$^{-1}$).

\begin{figure}
\includegraphics[angle=0,scale=0.3]{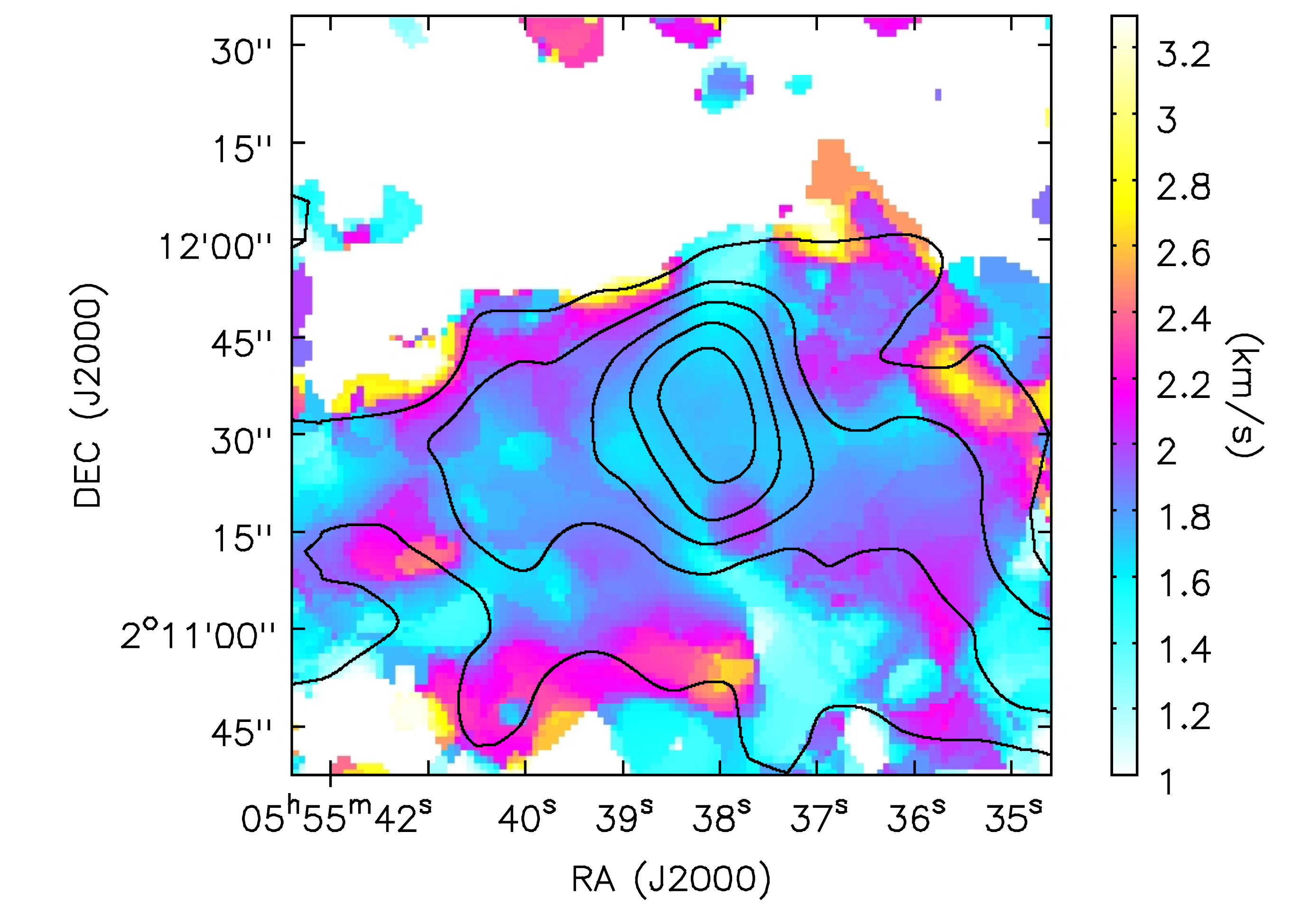}
\caption{The radial velocity (moment 1) map (color) toward G204.4-11.3 in N$_2$H$^+$
superimposed on the integrated intensity (moment 0) map.
The contour levels are
the same as those in Figure 12.
}
\end{figure}

\subsection{G207.3-19.8}

G207.3-19.8 is located in the Orion A GMC, and close to No. 19 CO emission peak identified by \citet{1986ApJ...303..375M}. G207N is associated with the Herbig-Haro object HH58, which is a Class 0 like source.  G207S is starless.
Figure 11 shows maps toward G207N.
The N$_2$H$^+$ emission is well correlated with the 850 $\mu$m distribution.
The N$_2$H$^+$ emission shows a peak toward G207N1 associated with HH58.
The HC$_3$N also shows a peak toward G207N1.
Weak emission is observed also toward G207N2, G207N3, G207N4, and G207N5 in N$_2$H$^+$, but not in HC$_3$N.
We have not detected either the 82 GHz or 94 GHz CCS emission.
In single pointing observation toward the T70 position, which was mistakenly set north of G207N, we have not detected either DNC or N$_2$D$^+$.

Figure 12 shows the N$_2$H$^+$ map toward G207S.
The N$_2$H$^+$ emission is more or less correlated with the 850 $\mu$m distribution,
but their peak positions do not coincide completely with each other.
We have not detected either the HC$_3$N, 82 GHz or 94 GHz CCS emission.

\subsection{G224.4-0.6}

This clump is located in 
L1658 \citet{1962ApJS....7....1L,2005PASJ...57S...1D}, 
the Orion Southern Filament \citep{1989ARAA..27...41G}, CMa R1 region \citep{1984PASJ...36..517N}, and  CMa OB1 \citep{bli78,1996JKASS..29..193K}
and near the dark cloud Khavtassi 330 \citep{kha60}.
Figure 13 shows maps toward G224.4-0.6.
The N$_2$H$^+$ emission is well correlated with the 850 $\mu$m distribution.  
Both tracers have two prominent intensity peaks, i.e., G224NE and G224S.  
We see slight peak position offsets.
The HC$_3$N  emission is not well correlated with the 850 $\mu$m distribution, and shows a stronger peak toward the southern 850 $\mu$m source.
The 82 GHz and 94 GHz CCS  emissions have local emission peaks between the two 850 $\mu$m sources.
T70 observations were carried out toward G224NE, and we detected DNC and HN$^{13}$C.
Figure 17 shows the moment 1 radial velocity map in N$_2$H$^+$.  We see a steep velocity gradient.  The SE-NW gradient is about 0.9 km s$^{-1}$ arcmin$^{-1}$ or 3 km s$^{-1}$ pc$^{-1}$.

\begin{figure}
\includegraphics[angle=0,scale=0.3]{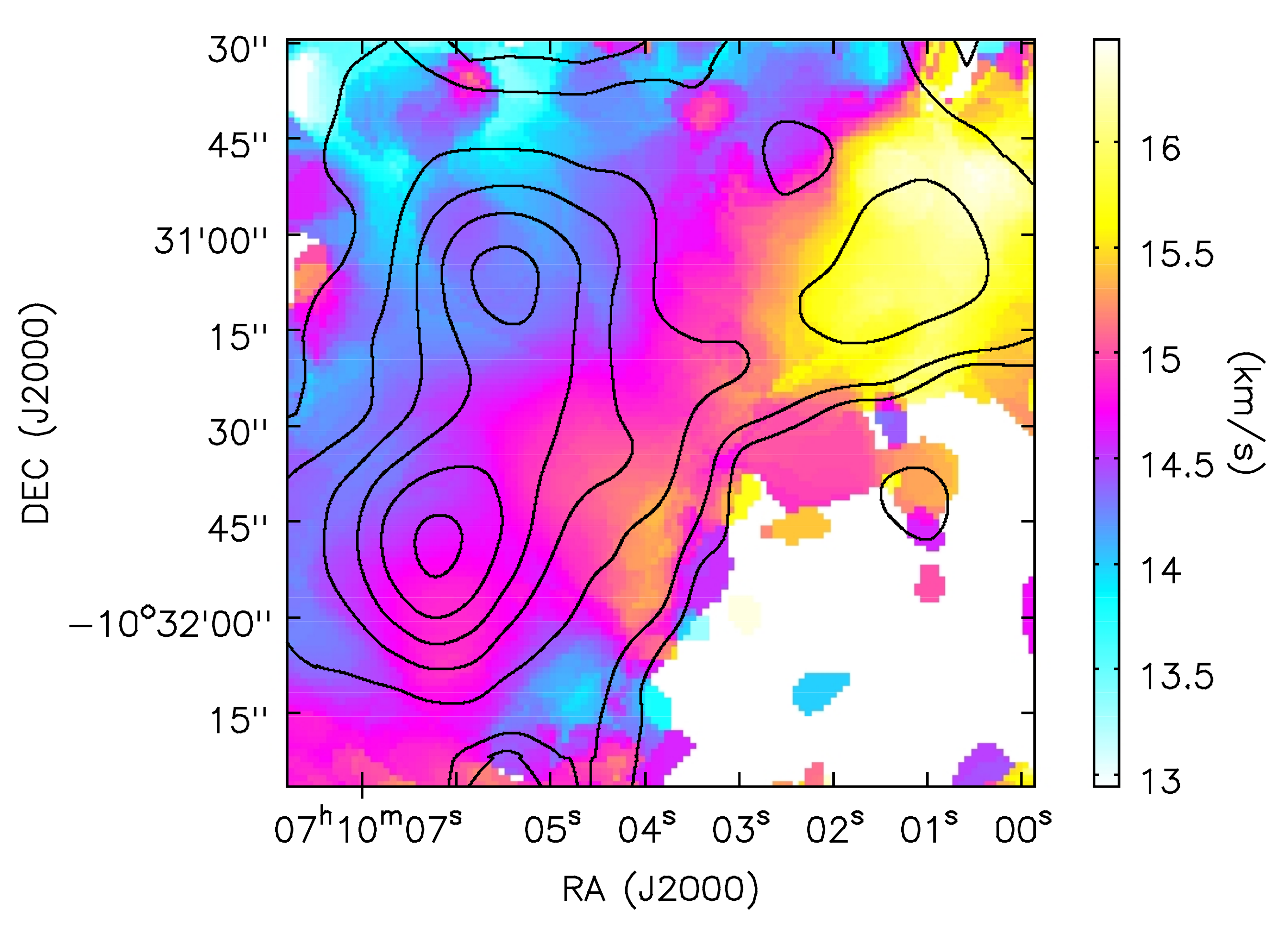}
\caption{The radial velocity (moment 1) map (color) toward G224.4-0.6 in N$_2$H$^+$
superimposed on the integrated intensity (moment 0) map.
The contour levels are
the same as those in Figure 16.}
\end{figure}

\section{Discussion}

\subsection{Column Density Ratios}

\citet{2006ApJ...646..258H} indicated the evolutionary sequence of starless cores by using column density ratios such as $N$(DNC)/$N$(HN$^{13}$C).
It seems that $N$(DNC)/$N$(HN$^{13}$C) in starless cores increases with core evolution ($<$0.66 to 3).
Using their data, we can show that $N$(N$_2$H$^+$)/$N$(CCS) is
$\lesssim$ 0.12 for young starless cores (L1495B, L1521B, L1521E, TMC-1, and L492).
According to \cite{2014PASJ...66...16T}, $N$(N$_2$H$^+$)/$N$(CCS) is usually $\lesssim 2-3$ for starless cores, but can reach $\sim$ 2$-$3 for evolved starless cores.
Star-forming cores generally give $N$(N$_2$H$^+$)/$N$(CCS) $\gtrsim 2-3$.

In the present observations, $N$(N$_2$H$^+$)/$N$(CCS) ranges from 0.4 to 3.7,
and
$N$(DNC)/$N$(HN$^{13}$C) ranges from 1.7 to 10.
These results suggest that our Planck cold clumps consist of various evolutionary stages including  relatively young starless cores and those on the verge of star formation.
In the present observations, $N$(N$_2$D$^+$)/$N$(N$_2$H$^+$) ranges from 0.1 to 1.4.
\cite{2006AA...460..709F} 
and
\cite{2011ApJ...743..196C}
studied the column density ratio of N$_2$D$^+$ to N$_2$H$^+$ (they call this the deuterium fractionation $D_{frac}$) toward massive protostellar cores, and compared it with those of low-mass prestellar cores by \cite{2005ApJ...619..379C}.  $N$(N$_2$D$^+$)/$N$(N$_2$H$^+$) is of order 10$^{-1}$ in low-mass prestellar cores \citep{2005ApJ...619..379C}, and of order 10$^{-2}$ in massive protostellar IRAS cores \citep{2006AA...460..709F}.  The ratio is estimated to be 0.35 and 0.08 at typical cold clouds, L134N and TMC-1N, respectively \citep{2000AA...356.1039T}.
Our N$_2$D$^+$ detected cores have larger $N$(N$_2$D$^+$)/$N$(N$_2$H$^+$) values than the massive protostellar IRAS cores of \citet{2006AA...460..709F}.

In G149.5-1.2 (contains a WISE source), G192N (contains a Class 0), G207N (contains HH58), and G207S, we did not detect the 82 GHz CCS emission over the OTF map regions.
The H$_2$ column density of our sources are 3$\times 10^{21}$ cm$^{-2}$ or higher (Tables 10 and 11).
In G149.5-1.2, the H$_2$ column density is lower than 1$\times 10^{22}$ cm$^{-2}$.
It is possible that a low column density is the reason for non-detection of CCS.
The H$_2$ column density toward G192N (contains Class 0) is as high as 6$\times 10^{22}$ cm$^{-2}$, so non-detection of CCS probably means that the gas is chemically evolved.
In G108S, the H$_2$ column density is as low as $\sim 1\times 10^{22}$ cm$^{-2}$, but we detected the 82 GHz CCS emission.
G089W (starless) should be chemically young, because $N$(N$_2$H$^+$)/$N$(CCS) is as small as 0.7
and $N$(N$_2$D$^+$)/$N$(N$_2$H$^+$) is as high as 1.4.
G157.6-12.2 (starless) is young, because  $N$(DNC)/$N$(HN$^{13}$C) is as high as 10.4.
G108N (starless) and G204NE (starless) show $N$(N$_2$H$^+$)/$N$(CCS) is $\sim$ 2, which is
close to the border between starless and star-forming cores in \cite{2014PASJ...66...16T}.

In the next subsection, we discuss the evolutionary stage of our Planck cold clumps on the basis of the column density ratios in Tables 8 and 9.  
We assume the filling factor is unity when we estimate the column density, but it is possible that the beam filling factor is less than unity.
If beam filling factors are similar between molecules, then the column density ratio will be less affected by unknown absolute values of the filling factor.
Because $T_k$ is $\lesssim$ 25 K in general, we can use $N$(N$_2$H$^+$)/$N$(CCS) as a chemical evolution tracer \citep{2014PASJ...66...16T}. The beam sizes for CCS and NH$_3$  differ by a factor of four, and $N$(NH$_3$)/$N$(CCS) suffers from sampling/averaging over very different size scales.  
We thus consider $N$(NH$_3$)/$N$(CCS) much less reliable. 

\subsection{Chemical Evolution Factor}

We introduce a new parameter to represent the chemical evolution by using molecular column density ratios,
the chemical evolution factor (CEF).
We define CEF so that starless cores have CEFs of $\sim$ -100 to 0, 
and star-forming cores show $\sim$ 0 to 100.
Starless cores having CEF $\sim$ 0 are regarded as being on the verge of star formation.
We use the form of 
CEF = log([$N$(A)/$N$(B)]/[$N_0$(A)/$N_0$(B)])*$d$
for the column density ratio of molecule A to molecule B.
$N_0$(A)/$N_0$(B) is chosen to be the column density ratio at the time of star formation.
For ($N$(N$_2$H$^+$)/$N$(CCS) and $N$(NH$_3$)/$N$(CCS),
$N_0$(A)/$N_0$(B) corresponds to the border between starless and star-forming cores.
For $N$(DNC)/$N$(HN$^{13}$C) and $N$(N$_2$D$^+$)/$N$(N$_2$H$^+$),
$N_0$(A)/$N_0$(B)
is the highest value observed for starless cores.
The factor $d$ is determined so that all starless cores range approximately from
-100 to 0.
By taking into account the observational results of
\citet{1992ApJ...392..551S,2005ApJ...619..379C,2006ApJ...646..258H,2014PASJ...66...16T},
we define CEF as CEF =
log($N$(N$_2$H$^+$)/$N$(CCS)/2.5)*50,
log($N$(DNC)/$N$(HN$^{13}$C)/3)*120,
log($N$(N$_2$D$^+$)/$N$(N$_2$H$^+$)/0.3)*50,
and
log($N$(NH$_3$)/$N$(CCS)/70)*70,
for  
starless cores with $T_k$ $\sim$ 10$-$ 20  K at a spatial resolution of order 0.015-0.05 pc (for 0.1-pc sized structure ``molecular cloud core'').
These expressions should be only valid for the above-mentioned temperature range and spatial resolution, because
we determined the CEF by using data obtained for molecular clumps or molecular cloud cores
having such temperatures and observed at such spatial resolutions.
The chemical reaction will depend on density, temperature, radiation strength, cosmic-ray strength, etc.
The deuterium fraction will be lower for warm cores
 \citep{1979ApJ...228..748S,1987IAUS..120..311W,1992AA...256..595S,2010PASJ...62.1473T}. 
It seems that the deuterium fraction decreases after the onset of star formation
\citep{2009A&A...496..731E,2012ApJ...747..140S,2014MNRAS.440..448F,2015ApJ...803...70S}.
Because the nature of this decrease has not yet been fully characterized observationally,
we do not use star-forming cores for CEF based on the deuterium fractionation.
Figure 18 shows the resulting CEF using the data in the literature 
\citep{2005ApJ...619..379C,2006ApJ...646..258H,2009ApJ...699..585H}.

Tables 12 and 13 and Figures 19 and 20 show the CEFs estimated in the present study.
We regard cores as star-forming if the first entry of comments in Table 3 suggests a possibility of star formation
(e.g., Class 0, Class I etc).
Our beam size with receivers TZ1 and T70 corresponds to 0.1 and 0.3 pc at a distance of 700 pc and 3.5 kpc, respectively.
To see the effect of very different spatial resolution (and probably very different volume density and very different beam-filling factor),
we show sources located beyond 1 kpc in parentheses.
In this paper we treat only starless cores for CEF, because evolution of star-forming cores has not well been characterized yet.
If we mistakenly identified star-forming cores as starless,
we may mistakenly obtain lower CEF due to decrease in the deuterium fractionation after star formation.
If this is the case, we may see inconsistency between CEF based on he deuterium fractionation
and that based on  $N$(N$_2$H$^+$)/$N$(CCS).
CEF(average) does not include the upper or lower limit to $N$(N$_2$H$^+$)/$N$(CCS).
For G108N, CEF(DNC,HN$^{13}$C) is a lower limit, but because this is a starless core, we will not provide a positive CEF(DNC,HN$^{13}$C) value.
That means CEF(DNC,HN$^{13}$C) = -9 to 0 nominally.  However, this does not mean the estimate is very accurate.
Then, we simply adopt this lower limit value of -9 to avoid being misleading.
Figures 19 and 20 show that
G174.0-15.8 is chemically
young, and 
G204NE seems to be in the intermediate stage (on the verge of star formation).

The advantage of Planck cold clumps is that they are cold ($\lesssim$ 20 K), and
less affected by temperature effects.
We have detected the N$_2$H$^+$ (1-0) emission from all the 13 clumps mapped with the Nobeyama 45 m telescope and receiver TZ1.
The optically thin critical density for this transition is 6.1$\times10^4$ and 4.1$\times10^4$~cm$^{-3}$ for 10 and 20 K, respectively \citep{2015PASP..127..299S}.
The volume density $n$ for dense cores in cold molecular clouds detected in N$_2$H$^+$ (1-0) is $\gtrsim$ 3$\times10^4$~cm$^{-3}$ \citep{2002ApJ...572..238C}.
Thus, we can assume that the 13 mapped Planck cold clumps have densities higher than $\gtrsim$ 3$\times10^4$~cm$^{-3}$.
With increasing density, the chemical reaction timescale will decrease.
Then, the change in CEF will correspond to longer timescales for lower densities.
The physical nature of the Planck cold clumps such as radius, mass, and volume density will be discussed in detail in a separate paper.

\begin{figure}
\includegraphics[angle=0,scale=0.5]{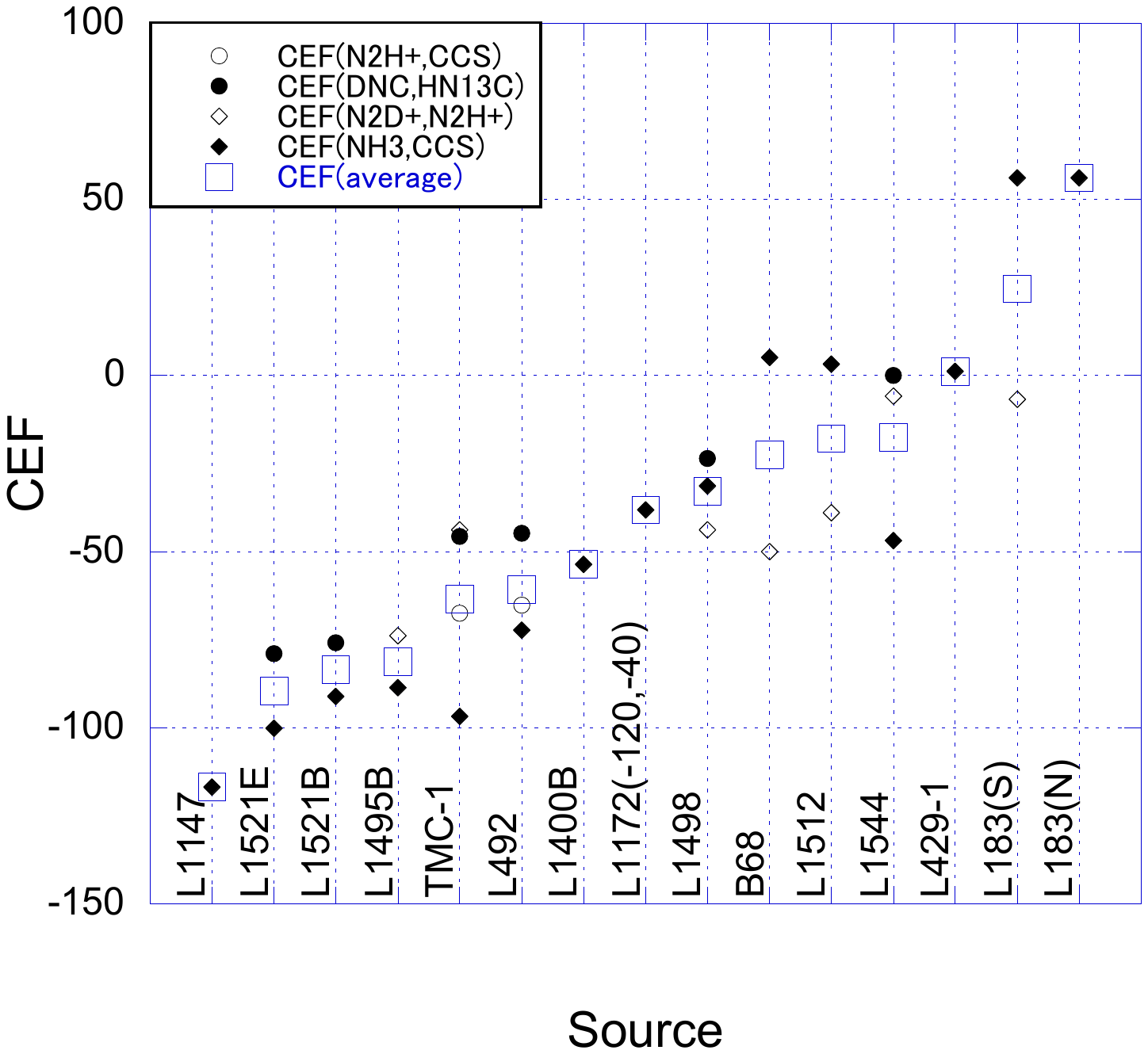}
\caption{Chemical Evolution Factor (CEF) for starless
sources in the literature.
}
\end{figure}

\begin{figure}
\includegraphics[angle=0,scale=0.5]{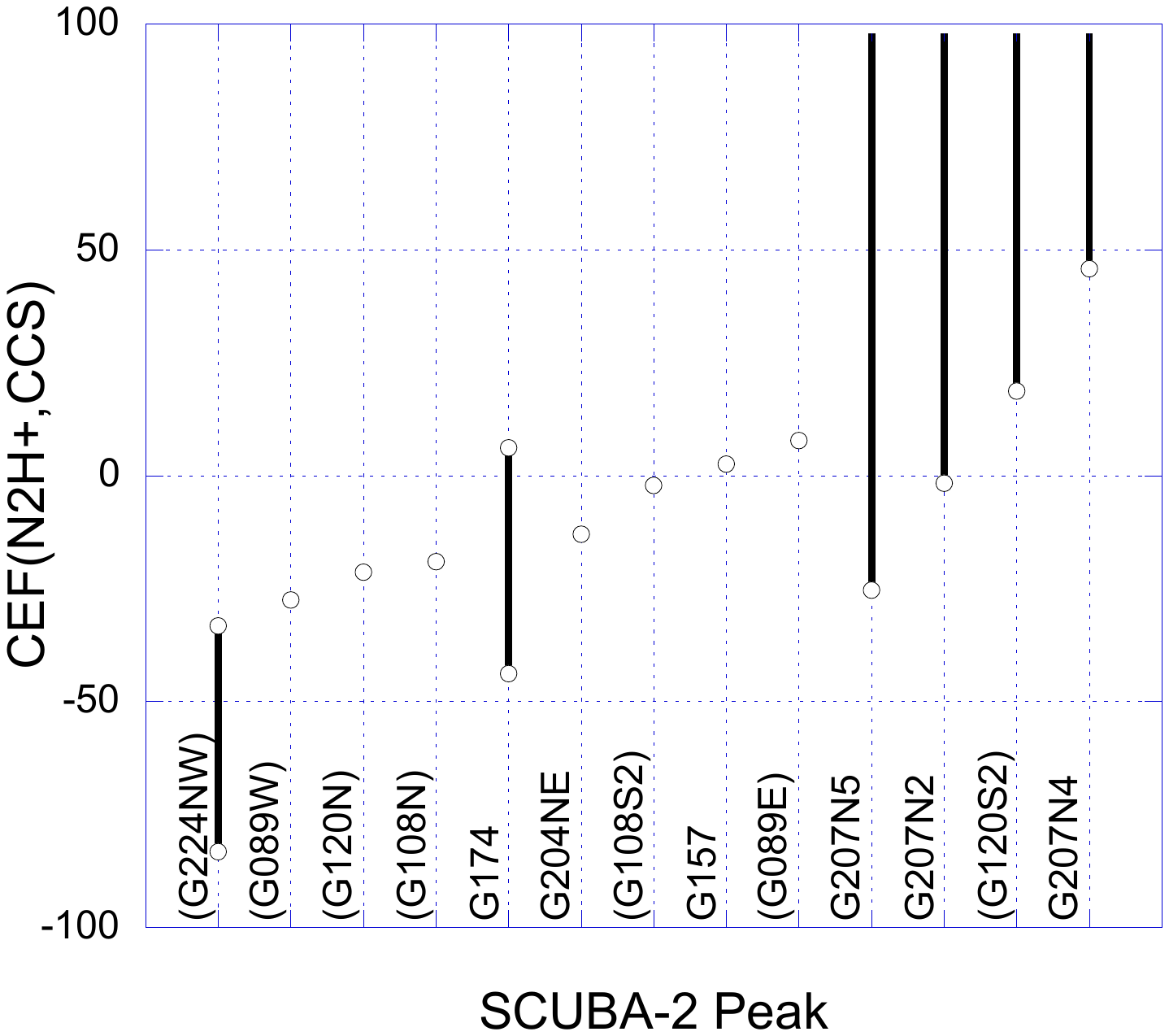}
\caption{Chemical Evolution Factor (CEF) for starless SCUBA-2 peaks 
based on the column density ratio of $N$(N$_2$H$^+$)/$N$(CCS).
The source name in parentheses means distance is larger than 1kpc.
The range reflects the uncertainty for the N$_2$H$^+$ optical depth for cores for which hyperfine fitting was not successful.}
\end{figure}

\begin{figure}
\includegraphics[angle=0,scale=0.5]{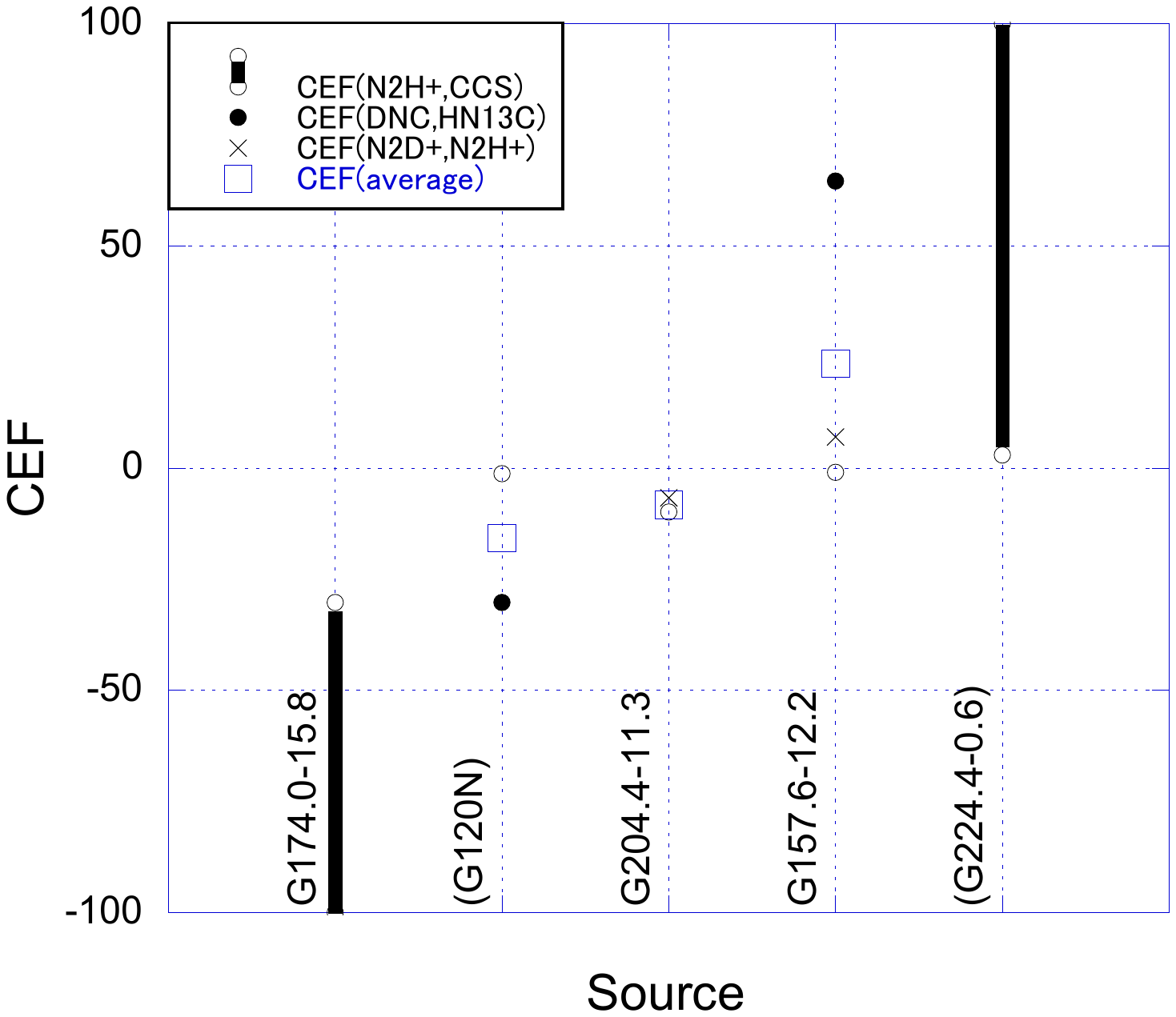}
\caption{Chemical Evolution Factor (CEF) at starless T70 positions based on multiple column density ratios.
The source name in parentheses means distance is larger than 1kpc.
The range reflects the upper and lower limit to $N$(N$_2$H$^+$)/$N$(CCS).}
\end{figure}

\subsection{Molecular Distribution}

Next, we investigate the morphology.  In G089.9-01.9, the 82 GHz and 94 GHz CCS emission (young molecular gas) is distributed as if it surrounds the N$_2$H$^+$ core (evolved gas). 
In G157.6-12.2, the 82 GHz CCS emission is distributed as if it surrounds the N$_2$H$^+$ core.
Such  configurations were previously reported in  \citet{2001ApJ...552..639A} for L1544, and also starless NH$_3$ core surrounded by CCS configurations are also observed
by \citet{2000ApJS..128..271L} for L1498 and \citet{2014ApJ...789...83T} for Orion A GMC. L1544 shows evidence of the prestellar
collapse.  
Therefore, these cores could be good targets for further studies for the initial conditions of star formation.
For  G157.6-12.2, CEF is $\gtrsim$ 0, and
its linewidth is as narrow as 0.3 km s$^{-1}$.
It is possible that this core is a coherent core that has largely dissipated turbulence, and is
on the verge of star formation 
\citep{2014ApJ...789...83T,2016MNRAS.459.4130O}.
In G120S1, G120S2, and G224.4-0.6 N$_2$H$^+$ and CCS distribution are largely different. 
G089.9-01.9, G120S, G157.6-12.2 are cold ($<$ 11K), which may suggest depletion of CCS (cf.,
\citet{2001ApJ...552..639A,2002ApJ...570L.101B,2002ApJ...572..238C}.
On the other hand, N$_2$H$^+$ peaks are detected in the HC$_3$N emission in these cores.
In G149N, the HC$_3$N emission (young molecular gas) is distributed on the both sides of the N$_2$H$^+$ core (evolved gas), which is also interesting.
In G108N and G108S, the 82 GHz GHz CCS and N$_2$H$^+$ emission coexist roughly, but, again, this could be due to poorer physical resolution or a projection effect.



\section{Summary}
Thirteen Planck cold clumps were observed with the James Clerk Maxwell Telescope/SCUBA-2 and with the Nobeyama 45 m radio telescope.  
The N$_2$H$^+$ spatial distribution is  similar to SCUBA-2 dust distribution.  
The spatial distribution of HC$_3$N, 82 GHz CCS, and 94 GHz CCS emission 
is often different from that of the N$_2$H$^+$ emission.  
The CCS emission is often very clumpy.  In G089.9-01.9 and G157.6-12.2, the CCS emission surrounds the
N$_2$H$^+$ core, which resembles the case of L1544 and suggests that they are on the verge of star formation.  
The detection rate of N$_2$D$^+$ is 50\%.  We investigated chemical evolutionary stages of starless Planck cold clumps
using the newly defined Chemical Evolution Factor (CEF). 
We found that G174.0-15.8 is chemically
young, and
G089E, G157.6-12.2, and G204NE seem in the intermediate stage (on the verge of star formation).
In addition, we observed NH$_3$, and determined the kinetic temperature $T_k$.

\clearpage

\floattable
\begin{deluxetable}{lllccc}
\tablecaption{Observed Lines \label{tab:tbl-1}}
\tablecolumns{6}
\tablenum{1}
\tablewidth{0pt}
\tablehead{
\colhead{Line} &
\colhead{Frequency} &
\colhead{Frequency Reference}&
\colhead{Upper Energy Level $E_u$}&
\colhead{Receiver} &
\colhead{Observing Mode}
}
\startdata
CCS $J_N$ = 7$_6-6_5$&81.505208 GHz&\citet{cum86}&15.3 K&TZ1&OTF	\\
CCS $J_N$ = 8$_7-6_5$&93.870107 GHz&\citet{1990ApJ...361..318Y}&19.9 K&TZ1&OTF	\\
HC$_3$N $J$ = 9$-$8&81.8814614 GHz&\citet{1998JQSRT..60..883P}&19.7 K&TZ1&OTF	\\
N$_2$H$^+$ $J$ = 1$-$0&93.1737767 GHz&\citet{1995ApJ...455L..77C}&4.5 K&TZ1&OTF	\\
DNC $J$ = 1$-$0&76.3057270 GHz&\citet{1998JQSRT..60..883P}&T70&3.7 K&single pointing\\
HN$^{13}$C $J$ = 1$-$0&87.090859 GHz&\citet{1979ApJ...232L..65F}&4.2 K&T70&single pointing\\
N$_2$D$^+$ $J$ = 1$-$0&77.1096100 GHz&\citet{1998JQSRT..60..883P}&3.7 K&T70&single pointing\\
cyclic C$_3$H$_2$ $J_{K_aK_c}$ = 2$_{12}-$1$_{01}$&85.338906 GHz&\citet{1981ApJ...246L..41T}&4.1 K&T70&single pointing\\
NH$_3$ $(J, K)$ = (1, 1)&23.694495 GHz&\citet{1983ARAA..21..239H}&23.4 K&H22&single pointing\\
\enddata
\end{deluxetable}


\floattable
\begin{deluxetable}{lcccccccccc}
\rotate
\tablecaption{Coordinates for the Observations \label{tab:tbl-2}}
\tablecolumns{11}
\tablenum{2}
\tablewidth{0pt}
\tablehead{
\colhead{Source} &
\colhead{OTF Reference Center} &
\colhead{}&
\colhead{OTF area} &
\colhead{T70} &
\colhead{}&
\colhead{H22} &
\colhead{}&
\colhead{OFF} &
\colhead{} &
\colhead{Distance}
\\
\colhead{}&
\colhead{RA(J2000.0)} &
\colhead{Dec(J2000.0)} &
\colhead{}&
\colhead{RA(J2000.0)} &
\colhead{Dec(J2000.0)} &
\colhead{RA(J2000.0)} &
\colhead{Dec(J2000.0)} &
\colhead{RA(J2000.0)} &
\colhead{Dec(J2000.0)} &
\colhead{(kpc)}
}
\startdata
G089.9-01.9	&	21:20:04.35    	&	46:54:46.8	&	$2\arcmin\times 2\arcmin$	&	21:20:04.35	&	46:54:46.8	&	21:20:04.35	&	46:54:46.8	&	21:19:20	&	46:54:00	&	0.62	\\
G108.8-00.8N (G108N)	&	22:58:56.2	&	58:58:21.7	&	$2\arcmin\times 2\arcmin$	&	22:58:53.6	&	58:58:21.7	&	22:58:56.2	&	58:58:21.7	&	22:58:15  	&	59:01:00	&	3.2	\\
G108.8-00.8S (G108S)	&	22:58:40.8	&	58:55:34.4	&	$2\arcmin\times 2\arcmin$	&		&		&	22:58:40.8	&	58:55:34.4	&	22:58:15	&	59:01:00	&	3.2	\\
G120.7+2.7N (G120N)	&	00:29:26.11	&	65:27:14.3	&	$2\arcmin\times 2\arcmin$	&	00:29:29.3	&	65:27:14.3	&	00:29:26.11	&	65:27:14.3	&	00:29:20	&	65:20:00	&	1.82	\\
G120.7+2.7S (G120S) 	&	00:29:46.76	&	65:25:47.8	&	$2\arcmin\times 2\arcmin$	&		&		&	00:29:46.76	&	65:25:47.8	&	00:29:20	&	65:20:00	&	1.82	\\
G149.5-1.2	&	03:56:53	&	51:48:00	&	$3\arcmin\times 3\arcmin$	&	03:56:57.3	&	51:49:00	&	03:56:57.3	&	51:49:00	&	03:56:15	&	51:44:00	&	0.84	\\
G157.6-12.2	&	03:51:53.6	&	38:15:22.6	&	$2\arcmin\times 2\arcmin$	&	03:51:53.6	&	38:15:22.6	&	03:51:53.6	&	38:15:22.6	&	03:51:20	&	38:10:00	&	0.45	\\
G174.0-15.8	&	04:32:45.30    	&	24:21:51.7	&	$2\arcmin\times 2\arcmin$	&		&		&	04:32:45.30	&	24:21:51.7	&	04:33:09.19	&	24:23:42.7	&	0.15	\\
G192.33-11.88	&	05:29:54.16	&	12:16:53.0	&	$2\arcmin\times 2\arcmin$	&	05:29:54.16	&	12:16:53.0	&	05:29:54.16	&	12:16:53.0	&	05:30:05	&	12:10:00	&	0.42	\\
G202.00+2.65	&	06:40:55.10	&	10:56:16.2	&		&		&		&	06:40:55.10	&	10:56:16.2	&	06:40:43.93	&	11:01:40.8	&	0.76	\\
G202.31-8.92	&	06:00:10	&	05:15:00	&		&		&		&	06:00:10.0	&	05:15:00.0	&	06:00:18	&	05:20:00	&	0.42	\\
G204.4-11.3	&	05:55:38.54    	&	02:11:35.6	&	$2\arcmin\times 2\arcmin$	&	05:55:38.54	&	02:11:35.6	&	05:55:38.54	&	02:11:35.6	&	05:55:48.98	&	02:05:37.0	&	0.42	\\
G207.3-19.8N (G207N)	&	05:30:48	&	-04:11:30	&	$4\arcmin\times 4\arcmin$	&	05:30:48.5	&	-04:09:40.0	&	05:30:46.5	&	-04:10:30	&	05:31:35	&	-04:12:22	&	0.42	\\
G207.3-19.8S (G207S)	&	05:31:03.00	&	-04:16:20	&	$4\arcmin\times 4\arcmin$	&		&		&	05:31:03.00	&	-04:17.05.3	&	05:31:35	&	-04:12:22	&	0.42	\\
G224.4-0.6	&	07:10:03.83	&	-10:31:28.21	&	$2\arcmin\times 2\arcmin$	&	07:09:59.8	&	-10:31:18.2	&	07:10:01.12	&	-10:30:58.2	&	07:09:55	&	-10:28:00	&	1.1	\\
\enddata
\end{deluxetable}


\begin{deluxetable*}{ccccc}
\tablenum{3}
\tabletypesize{\scriptsize} \tablecolumns{5} \tablewidth{0pc}
\tablecaption{Parameters of SCUBA-2 Peaks} 
\tablehead{
\colhead{Source} & \colhead{SCUBA-2 peak} & \colhead{RA }  & \colhead{DEC } & comments\\
               &                   &  \colhead{(J2000)} & \colhead{ (J2000)}  &      } 
\startdata
G089.9-01.9   & G089E   & 21:20:08.5 & 46:54:52.4 & starless\\
              & G089W   & 21:20:04.4 & 46:54:47.4 & starless? an infrared source IRAS 21182+4641 on the east \\
G108.8-00.8   & G108N   & 22:58:57.5 & 58:58:30.2 & starless\\
              & G108S1  & 22:58:40.8 & 58:55:02.0 & starless\\
              & G108S2  & 22:58:41.1 & 58:55:37.1 & starless\\
              & G108S3  & 22:58:38.9 & 58:55:48.1 & Class I?\\
              & G108S4  & 22:58:34.2 & 58:55:48.2 & Class I?\\
G120.7+2.7    & G120S1  & 00:29:43.8 & 65:26:03.1 & WISE faint source only detected at 3.4 and 4.6 micron, \\
              &         &            &             & no AKARI source, a foreground star?\\
              & G120S2  & 00:29:38.5 & 65:26:17.8 & starless\\
              & G120N   & 00:29:29.1 & 65:27:18.4 & starless, offset from infrared sources\\
G149.5-1.2    & G149N   & 03:56:57.2 & 51:48:50.0 & WISE, class I? \\
              & G149S   & 03:56:50.4 & 51:46:57.4 & starless \\
G157.6-12.1   & G157    & 03:51:53.3 & 38:15:24.3 & starless \\
G174.0-15.8   & G174    & 04:32:44.9 & 24:21:39.6 & starless \\
G192.33-11.88 & G192N   & 05:29:54.5 & 12:16:55.0 & Class 0 \\
              & G192S   & 05:29:54.7 & 12:16:30.6 & proto-brown dwarf candidate \\
G202.00+2.65  & G202.0N & 06:40:56.1 & 10:56:34.9 & Class 0? Spitzer, Akari, WISE\\
              & G202.0M & 06:40:55.1 & 10:56:16.4 & starless\\
              & G202.0S & 06:40:54.8 & 10:55:40.1 & Class 0? Spitzer, WISE\\
G202.31-8.92  & G202.3  & 06:00:08.8 & 05:14:59.6 & starless\\
G204.4-11.3   & G204NE  & 05:55:38.4 & 02:11:35.5 & starless\\
              & G204SW  & 05:55:35.6 & 02:11:01.6 & class 0?\\
G207N  & G207N1  & 05:30:51.0 & -04:10:36.7 & class 0? HH 58? but SCUBA-2 core peak offset from infrared source\\
              & G207N2  & 05:30:50.8 & -04:10:14.3 & starless\\
              & G207N3  & 05:30:46.5 & -04:10:29.0 & starless\\
              & G207N4  & 05:30:44.7 & -04:10:27.4 & starless\\
              & G207N5  & 05:30:47.1 & -04:12:31.3 & starless\\
G207S  & G207S1  & 05:31:02.0 & -04:14:56.6 & starless\\
              & G207S2  & 05:31:03.7 & -04:15:48.3 & starless\\
              & G207S3  & 05:31:00.3 & -04:15:43.6 & starless\\
              & G207S4  & 05:31:03.2 & -04:17:00.3 & starless\\
              & G207S5  & 05:31:03.8 & -04:17:37.0 & starless\\
G224.4-0.6   & G224S   & 07:10:06.2 & -10:32:00.3 & IRAS 07077-1026, Akari, WISE, Spitzer\\
              & G224NE  & 07:10:05.6 & -10:31:11.8 & Akari, WISE, Spitzer\\
              & G224NW  & 07:10:00.8 & -10:30:58.2 & starless\\
\enddata
\tablenotetext{a}{Convolved with a beam of 18.8 arcsec}
\end{deluxetable*}


\floattable
\begin{deluxetable}{lccccccccccccccc}
\rotate
\tablecaption{Intensities
\tablenotemark{a}  
Observed with the Receiver TZ1 toward the SCUBA-2 Peaks\label{tbl-4}}
\tablenum{4}
\tablecolumns{16}
\tablewidth{0pt}
\tablehead{
\colhead{SCUBA-2 peak} &
\colhead{82 GHz CCS} &
\colhead{} &
\colhead{} &
\colhead{94 GHz CCS}&
\colhead{} &
\colhead{} &
\colhead{HC$_3$N} &
\colhead{} &
\colhead{} &
\colhead{N$_2$H$^+$} &
\colhead{} &
\colhead{} &
\colhead{} &
\colhead{} &
\colhead{}\\
\colhead{} &
\colhead{$T_A^*$} &
\colhead{$V_{LSR}$} &
\colhead{$\Delta v$}&
\colhead{$T_A^*$} &
\colhead{$V_{LSR}$} &
\colhead{$\Delta v$} &
\colhead{$T_A^*$} &
\colhead{$V_{LSR}$} &
\colhead{$\Delta v$} &
\colhead{$T_A^*$} &
\colhead{$V_{LSR}$} &
\colhead{$\Delta v$} &
\colhead{$T_{ex}$} &
\colhead{$\tau$(main)} &
\colhead{Integrated Intensity}\\
\colhead{}&
\colhead{(K)} &
\colhead{(km s$^{-1}$)}&
\colhead{(km s$^{-1}$)}&
\colhead{(K)} &
\colhead{(km s$^{-1}$)}&
\colhead{(km s$^{-1}$)}	&
\colhead{(K)} &
\colhead{(km s$^{-1}$)}&
\colhead{(km s$^{-1}$)}&
\colhead{(K)}&
\colhead{(km s$^{-1}$)}&
\colhead{(km s$^{-1}$)}&
\colhead{}&
\colhead{}&
\colhead{(K km s$^{-1}$)} 
}
\startdata
G089E	&	0.60 			&	1.94			&	0.14 			&	0.27 			&	1.80 			&	0.68 			&	0.56 			&	1.82			&	0.49 			&	0.8 				&	1.82 			&	0.40 			&	4.6 	$\pm$	0.4 	&	6.2 	$\pm$	2.0 	&		\\
G089W	&	0.51 			&	1.31			&	0.33 			&	0.43 			&	1.43 			&	0.42 			&	1.08 			&	1.29			&	0.47 			&	1.1 				&	1.50 			&	0.63 			&	8.7 	$\pm$	4.2 	&	1.4 	$\pm$	1.2 	&		\\
G108N	&	0.18			&	-49.38 			&	1.53 			&		$<$	0.16 	&				&				&				&				&				&	0.5 				&	-49.40 			&	1.20 			&	5.8 	$\pm$	3.8 	&	0.9 	$\pm$	1.3 	&		\\
G108S1	&	0.19 			&	-49.00 			&	0.78 			&				&				&				&	0.28 			&	-48.84			&	0.33 			&			$<$	0.10 	&				&				&				&				&		\\
G108S2	&	0.18 			&	-49.58 			&	1.44 			&				&				&				&	0.22 			&	-49.66			&	1.39 			&	0.3 				&	-49.78 			&	1.30 			&	3.4 	$\pm$	0.2 	&	3.2 	$\pm$	1.5 	&		\\
G108S3	&		$<$	0.12 	&				&				&				&				&				&				&				&				&	0.2 				&	$\sim$-50			&				&				&				&	0.5 	\\
G108S4	&		$<$	0.12 	&				&				&				&				&				&				&				&				&	0.3 				&	-51.20 			&	1.22 			&	3.8 	$\pm$	0.8 	&	1.9 	$\pm$	1.9 	&		\\
G120S1	&	0.25 			&	-18.01 			&	1.48 			&	0.21			&	-18.12 			&	1.48 			&	0.51 			&	-18.22			&	1.27 			&	0.6 				&	-18.33 			&	1.33 			&	4.8 	$\pm$	0.6 	&	1.9 	$\pm$	0.7 	&		\\
G120S2	&		$<$	0.14 	&				&				&		$<$	0.15 	&				&				&	0.79 			&	-18.16			&	0.92 			&	0.4 				&	-18.06 			&	0.98 			&	3.9 	$\pm$	0.3 	&	4.0 	$\pm$	2.0 	&		\\
G120N	&	0.21 			&	-18.44 			&	0.95 			&		$<$	0.17 	&				&				&	0.39 			&	-18.57			&	1.20 			&	1.2 				&	-18.46 			&	1.24 			&	22.7 	$\pm$	29.0 	&	0.3 	$\pm$	0.5 	&	2.6 	\\
G149N	&		$<$	0.16 	&				&				&		$<$	0.19 	&				&				&				&				&				&	0.6 				&	-7.51 			&	0.41 			&	4.5 	$\pm$	1.3 	&	3.5 	$\pm$	3.3 	&		\\
G149S	&		$<$	0.16 	&				&				&		$<$	0.19 	&				&				&				&				&				&			$<$	0.17 	&				&				&				&				&		\\
G157	&	0.57 			&	-7.62 			&	0.28 			&		$<$	0.19 	&				&				&				&				&				&	0.8 				&	-7.63 			&	0.29 			&	4.2 	$\pm$	0.1 	&	16.5 	$\pm$	5.0 	&		\\
G174	&	0.55 			&	6.27 			&	0.33 			&		$<$	0.16 	&				&				&				&				&				&	0.7 				&	6$-$7			&				&				&				&	0.5 	\\
G192N	&		$<$	0.16 	&				&				&		$<$	0.17 	&				&				&				&				&				&	0.6 				&	$\sim$6			&				&				&				&	1.4 	\\
G192S	&		$<$	0.16 	&				&				&		$<$	0.17 	&				&				&	0.37 			&	12.17			&	0.75 			&	1.3 				&	12.23 			&	0.55 			&	5.7 	$\pm$	0.2 	&	6.2 	$\pm$	0.9 	&		\\
G204NE	&	0.38 			&	1.74 			&	0.71 			&		$<$	0.17 	&				&				&				&				&				&	1.6 				&	1.57 			&	0.47 			&	6.0 	$\pm$	0.4 	&	7.5 	$\pm$	1.6 	&		\\
G204SW	&		$<$	0.21		&				&				&		$<$	0.20 	&				&				&				&				&				&	0.5 				&	1.74 			&	0.76 			&	5.2 	$\pm$	3.1 	&	1.8 	$\pm$	2.9 	&	1.0 	\\
G207N1	&		$<$	0.22 	&				&				&		$<$	0.22 	&				&				&	0.44 			&	10.48			&	1.20 			&	0.9 				&	10.71 			&	1.10 			&	6.1 	$\pm$	1.8 	&	2.4 	$\pm$	1.8 	&		\\
G207N2	&		$<$	0.22 	&				&				&		$<$	0.22 	&				&				&				&				&				&	0.8 				&	11.20 			&	0.45 			&	5.6 	$\pm$	1.6 	&	3.6 	$\pm$	2.8 	&		\\
G207N3	&		$<$	0.22 	&				&				&		$<$	0.22 	&				&				&				&				&				&					&				&				&				&				&		\\
G207N4	&		$<$	0.22 	&				&				&		$<$	0.22 	&				&				&				&				&				&	0.4 				&	11.28 			&	0.54 			&	3.5 	$\pm$	0.1 	&	21.5 	$\pm$	12.4 	&	0.9 	\\
G207N5	&		$<$	0.22 	&				&				&		$<$	0.22 	&				&				&				&				&				&	0.9 				&	$\sim$11			&				&				&				&	0.7 	\\
G224S	&	0.22 			&	14.44 			&	2.59 			&		$<$	0.19 	&				&				&	0.49 			&	14.4			&	2.17 			&	1.5 				&	$\sim$15			&				&				&				&	3.9 	\\
G224NE	&	0.15 			&	13.87 			&	3.46 			&		$<$	0.19 	&				&				&	0.36 			&	13.75			&	1.46 			&	1.6 				&	13.95 			&	1.46 			&	13.6 	$\pm$	4.2 	&	0.9 	$\pm$	0.4 	&		\\
G224NW	&	0.23 			&	16.21 			&	3.04 			&		$<$	0.19 	&				&				&	0.74 			&	16.38			&	2.00 			&	0.6 				&	$\sim$16			&				&				&				&	0.2 	\\
\enddata
\tablenotetext{a}{The upper limit to the intensity is defined as 3$\sigma$, where $\sigma$ is the rms noise level at 1 km s$^{-1}$ bin.}
\end{deluxetable}


\floattable
\begin{deluxetable}{lcccccccccccccc}
\rotate
\tablecaption{Intensities
\tablenotemark{a}  
Observed with the Receiver TZ1 toward T70 Position\label{tbl-5}}
\tablenum{5}
\tablecolumns{15}
\tablewidth{0pt}
\tablehead{
\colhead{Source} &
\colhead{82 GHz CCS} &
\colhead{} &
\colhead{} &
\colhead{94 GHz CCS}&
\colhead{} &
\colhead{} &
\colhead{HC$_3$N} &
\colhead{} &
\colhead{} &
\colhead{N$_2$H$^+$} &
\colhead{} &
\colhead{} &
\colhead{} &
\colhead{}\\
\colhead{} &
\colhead{$T_A^*$} &
\colhead{$V_{LSR}$} &
\colhead{$\Delta v$}&
\colhead{$T_A^*$} &
\colhead{$V_{LSR}$} &
\colhead{$\Delta v$} &
\colhead{$T_A^*$} &
\colhead{$V_{LSR}$} &
\colhead{$\Delta v$} &
\colhead{$T_A^*$} &
\colhead{$V_{LSR}$} &
\colhead{$\Delta v$} &
\colhead{$T_{ex}$} &
\colhead{$\tau$(main)}\\
\colhead{}&
\colhead{(K)} &
\colhead{(km s$^{-1}$)}&
\colhead{(km s$^{-1}$)}&
\colhead{(K)} &
\colhead{(km s$^{-1}$)}&
\colhead{(km s$^{-1}$)}	&
\colhead{(K)} &
\colhead{(km s$^{-1}$)}&
\colhead{(km s$^{-1}$)}&
\colhead{(K)}&
\colhead{(km s$^{-1}$)}&
\colhead{(km s$^{-1}$)}&
\colhead{(K)}&
\colhead{}
}
\startdata
G089.9-01.9	&	0.53 			&	1.32			&	0.32			&	0.44 			&	1.44			&	0.47			&	1.23 			&	1.29			&	0.45			&	1.1 				&	1.49 			&	0.60 			&	9.0 	$\pm$	4.9 	&	1.4 	$\pm$	1.3 			\\
G108N	&	0.16 			&	-49.62			&	1.56			&		$<$	0.16 	&				&				&		$<$	0.12 	&				&				&	0.5 				&	-49.89 			&	1.39 			&				&						\\
G120N	&	0.21 			&	-18.44			&	0.95			&		$<$	0.17 	&				&				&	0.45 			&	-18.65			&	0.81			&	1.2 				&	-18.57 			&	1.13 			&	8.1 	$\pm$	1.5 	&	1.6 	$\pm$	0.6 			\\
G120S 	&	0.35 			&	-18.42			&	0.73			&	0.33 			&	-18.42			&	0.31			&	0.80 			&	-18.47			&	0.66			&			$<$	0.14 	&				&				&				&						\\
G149.5-1.2	&		$<$	0.16 	&				&				&		$<$	0.19 	&				&				&		$<$	0.14 	&				&				&	0.6 				&	-7.51 			&	0.41 			&	4.5 	$\pm$	1.3 	&	3.5 	$\pm$	3.3 			\\
G157.6-12.2	&	0.59 			&	-7.62			&	0.27			&		$<$	0.20 	&				&				&		$<$	0.15 	&				&				&	0.9 				&	-7.61 			&	0.29 			&	4.5 	$\pm$	0.2 	&	9.4 	$\pm$	1.9 			\\
G174.0-15.8	&	0.64 			&	6.27			&	0.44			&		$<$	0.16 	&				&				&		$<$	0.11 	&				&				&			$<$	0.15 	&				&				&				&						\\
G192.33-11.88	&		$<$	0.16 	&				&				&		$<$	0.17 	&	13.48			&	0.37			&		$<$	0.15 	&				&				&	0.7 				&	12.12 			&	1.13 			&	4.9 	$\pm$	0.3 	&	2.7 	$\pm$	0.6 			\\
G204.4-11.3	&	0.38 			&	1.74			&	0.71			&		$<$	0.20 	&				&				&		$<$	0.18 	&				&				&	1.4 				&	1.59 			&	0.47 			&	5.6 	$\pm$	0.2 	&	8.6 	$\pm$	1.5 			\\
G207N\tablenotemark{b}		&		$<$	0.22 	&				&				&		$<$	0.24 	&				&				&		$<$	0.19 	&				&				&			$<$	0.23 	&				&				&				&						\\
G224.4-0.6	&		$<$	0.18 	&				&				&		$<$	0.19 	&				&				&		$<$	0.17 	&				&				&	1.6 				&	13.95 			&	1.46 			&	13.6 	$\pm$	4.2 	&	0.9 	$\pm$	0.4 	\\		
\enddata
\tablenotetext{a}{The upper limit to the intensity is defined as 3$\sigma$, where $\sigma$ is the rms noise level at 1 km s$^{-1}$ bin.}
\tablenotetext{b}{incorrect position}
\end{deluxetable}


\floattable
\begin{deluxetable}{lcccccccccccccc}
\rotate
\tablecaption{Intensities
\footnote{.} 
Observed with the Receiver T70 \label{tbl-6}}
\tablenum{6}
\tablecolumns{15}
\tablewidth{0pt}
\tablehead{
\colhead{Source} &
\colhead{DNC}   &
\colhead{} &
\colhead{}   &
\colhead{HN$^{13}$C}  &
\colhead{} &
\colhead{}   &
\colhead{N$_2$D$^+$} &
\colhead{} &
\colhead{} &
\colhead{} &
\colhead{} &
\colhead{c-C$_3$H$_2$}   &
\colhead{} &
\colhead{}\\
\colhead{} &
\colhead{$T_A^*$}   &
\colhead{$V_{LSR}$} &
\colhead{$\Delta v$} &
\colhead{$T_A^*$}   &
\colhead{$V_{LSR}$} &
\colhead{$\Delta v$} &
\colhead{$T_A^*$}   &
\colhead{$V_{LSR}$} &
\colhead{$\Delta v$} &
\colhead{$T_{ex}$} &
\colhead{$\tau$(tau)} &
\colhead{$T_A^*$} &
\colhead{$V_{LSR}$} &
\colhead{$\Delta v$}\\
\colhead{}&
\colhead{(K)} &
\colhead{(km s$^{-1}$)}	&
\colhead{(km s$^{-1}$)}&
\colhead{(K)} &
\colhead{(km s$^{-1}$)}&
\colhead{(km s$^{-1}$)}	&
\colhead{(K)} &
\colhead{(km s$^{-1}$)}&
\colhead{(km s$^{-1}$)}&
\colhead{(K)}	&
\colhead{ }	&
\colhead{(K)}&
\colhead{(km s$^{-1}$)}&
\colhead{(km s$^{-1}$)}
}
\startdata
G089.9-01.9	&	0.90 &	1.49 	&	1.19 	&	0.44 			&	1.49 	&	0.79	&	0.23	&	1.50 	&	0.50 	&	3.7 	$\pm$	0.4 	&	2.09 	$\pm$	1.08 	&	0.69 &	1.39	&	0.78	\\
G108N	&	0.18 			&	-49.46 	&	2.48 	&		$<$	0.089 	&		&	2.48	&		$<$	0.036 	&		&		&				&				&	0.21			&	-49.84	&	0.95	\\
G120N	&	0.30 			&	-18.52 	&	1.51 	&	0.25 			&	-18.46 	&	1.353	&	0.05	$<$	0.036 	&		&		&				&				&	0.51			&	-18.45	&	1.41	\\
G149.5-1.2	&	0.17 			&	-7.47 	&	1.00 	&	0.12 			&	-7.31 	&	0.592	&	0.05	$<$	0.036 	&		&		&				&				&	0.28			&	-7.44	&	0.46	\\
G157.6-12.2	&	1.34 	&	-7.66 	&	0.90 	&	0.53 	&	-7.55 	&	0.564	&	0.45	&	-7.65 	&	0.31 	&	4.1 	$\pm$	0.2 	&	3.42 	$\pm$	0.66 	&	1.25			&	-7.61	&	0.35	\\
G174.0-15.8	&				&		&		&				&		&		&				&		&		&				&				&				&		&		\\
G192.33-11.88	&	0.81 	&	12.14 	&	1.06 	&	0.22 			&	12.32 	&	0.959	&	0.28		&	12.16 	&	0.59 	&	5.8 $\pm$	2.0 &	0.58 $\pm$	0.40 	&	0.42	&	12.19	&	0.84	\\
G204.4-11.3	&	1.69 &	1.64 	&	1.03 	&	0.74 	&	1.70 	&	0.79	&	0.38			&	1.63 	&	0.52 	&	4.7 	$\pm$	1.4 	&	1.68 	$\pm$	1.33 	&	1.47			&	1.64	&	0.66	\\
G207N\tablenotemark{b}		&		$<$	0.040 	&		&		&		$<$	0.026 	&		&		&		$<$	0.036 	&		&		&				&				&	0.05			&	11.12	&	1.64	\\
G224.4-0.6	&	0.23 			&	15.76 	&	1.42 	&	0.17 			&	15.73 	&	0.479	&		$<$	0.044 	&		&		&				&				&	0.19			&	15.69	&	1.35\\	
\enddata
\tablenotetext{a}{The upper limit to the intensity is defined as 3$\sigma$, where $\sigma$ is the rms noise level at 1 km s$^{-1}$ bin.}
\tablenotetext{b}{incorrect position}
\end{deluxetable}


\floattable
\begin{deluxetable}{lcccccc}
\rotate
\tablecaption{NH$_3$ Intensities 
\tablenotemark{a}  
Observed with the Receiver H22 and Derived Rotation and Kinetic Temperatures\label{tbl-7}}
\tablenum{7}
\tablecolumns{7}
\tablewidth{0pt}
\tablehead{
\colhead{Source} &
\colhead{NH$_3$ (1,1)}   &
\colhead{} &
\colhead{}   &
\colhead{NH$_3$ (2,2)}  &
\colhead{$T_{rot}$} &
\colhead{$T_k$}\\
\colhead{} &
\colhead{$T_A^*$}   &
\colhead{$V_{LSR}$} &
\colhead{$\Delta v$} &
\colhead{$T_A^*$}   &
\colhead{} &
\colhead{}\\
\colhead{}&
\colhead{(K)} &
\colhead{(km s$^{-1}$)}&
\colhead{(km s$^{-1}$)}&
\colhead{(K)} &
\colhead{(K)}&
\colhead{(K)}	
}
\startdata
G089.9-01.9	&	0.87 	&	1.59 	&	0.86 	&		0.11 	&		10.5	$\pm$	0.9	&		11.0 	$\pm$	1.1 	\\
G108N	&	0.49 	&	-49.56 	&	1.65 	&		0.10 	&		13.1	$\pm$	2.8	&		14.3 	$\pm$	3.6 	\\
G108S	&	0.22 	&	-49.82 	&	2.06 	&	$<$	0.03 	&	$<$	16.5			&	$<$	19.1 			\\
G120N	&	0.42 	&	-18.39 	&	1.61 	&		0.11 	&		13.1	$\pm$	4.1	&		14.3 	$\pm$	5.4 	\\
G120S 	&	0.57 	&	-18.47 	&	1.11 	&		0.15 	&		11.3	$\pm$	1.5	&		12.1 	$\pm$	1.9 	\\
G149.5-1.2	&	0.35 	&	-7.49 	&	0.96 	&	$<$	0.03 	&	$<$	13.4			&	$<$	14.8 			\\
G157.6-12.2	&	0.71 	&	-7.68 	&	0.72 	&	$<$	0.03 	&	$<$	10.7			&	$<$	11.4 			\\
G174.0-15.8	&	0.99 	&	6.19 	&	0.81 	&	$<$	0.03 	&	$<$+	10.5			&	$<$	11.1 			\\
G192.33-11.88	&	1.01 	&	12.06 	&	0.90 	&		0.21 	&		11.6	$\pm$	1.7	&		12.5 	$\pm$	2.1 	\\
G202.00+2.65	&	0.80 	&	5.08 	&	0.77 	&	$<$	0.06 	&	$<$	9.8			&	$<$	10.2 			\\
G202.31-8.92	&	0.60 	&	11.92 	&	0.89 	&	$<$	0.03 	&	$<$	10.9			&	$<$	11.6 			\\
G204.4-11.3	&	1.34 	&	1.56 	&	0.73 	&		0.14 	&		9.8	$\pm$	0.7	&		10.2 	$\pm$	0.8 	\\
G207N	&	0.87 	&	11.04 	&	1.04 	&		0.14 	&		12.9	$\pm$	5.6	&		14.1 	$\pm$	7.3 	\\
G207S	&	0.62 	&	11.58 	&	1.31 	&	$<$	0.04 	&		11.1	$\pm$	5.3	&		11.8 	$\pm$	6.5 	\\
G224.4-0.6	&	0.37 	&	14.99 	&	2.72 	&		0.11 	&		14	$\pm$	13	&		15.5 	$\pm$	17.7 	\\
\enddata
\end{deluxetable}


\floattable
\begin{deluxetable}{lccc}
\rotate
\tablecaption{Column Densities and Their Ratios toward SCUBA-2 Peak \label{tbl-8}}
\tablenum{8}
\tablecolumns{4}
\tablewidth{0pt}
\tablehead{
\colhead{SCUBA-2 peak} &
\colhead{$N$(CCS)}   &
\colhead{$N$(N$_2$H$^+$)} &
\colhead{$N$(N$_2$H$^+$)/$N$(CCS)}  \\
\colhead{} &
\colhead{(cm$^{-2}$)}   &
\colhead{(cm$^{-2}$)} &
\colhead{}
}
\startdata
G089E	&	3.5E+12			&	1.2E+13				&		3.6 						\\
G089W	&	6.4E+12			&	4.5E+12				&						0.7 						\\
G108N	&	5.2E+12			&	5.4E+12				&				1.0 						\\
G108S1	&	6.4E+12			&					&			\\
G108S2	&	1.1E+13			&	2.5E+13				&			2.3 						\\
G108S3	&		$<$	4.9E+12	&	1.7E+12	$-$	1.7E+13		&									$>$	0.3 	\\
G108S4	&		$<$	4.9E+12	&	1.3E+13				&										$>$	2.6 	\\
G120S1	&	9.8E+12			&	1.3E+13				&						1.3 						\\
G120S2	&		$<$	3.5E+12	&	2.1E+13				&									$>$	5.9 	\\
G120N	&	3.8E+12			&	3.5E+12				&								0.9 						\\
G149N	&		$<$	6.8E+12	&	7.2E+12				&												$>$	1.1 	\\
G149S	&		$<$	6.8E+12	&					&																\\
G157	&	8.9E+12			&	2.5E+13				&											2.8 						\\
G174	&	5.0E+12			&	1.7E+12	$-$	1.7E+13		&										0.3 	$-$	3.3 				\\
G192N	&		$<$	3.9E+12	&	4.6E+12	$-$	4.6E+13		&													$>$	1.2 	\\
G192S	&		$<$	6.9E+12	&	1.7E+13				&													$>$	2.4 	\\
G204NE	&	1.2E+13			&	1.7E+13				&									1.4 						\\
G204SW	&				&	6.7E+12				&							\\
G207N1	&		$<$	3.3E+12	&	1.2E+13				&										$>$	3.8 	\\
G207N2	&		$<$	3.3E+12	&	7.7E+12				&											$>$	2.3 	\\
G207N3	&		$<$	3.3E+12	&					&																\\
G207N4	&		$<$	3.3E+12	&	6.8E+13				&									$>$	21 	\\
G207N5	&		$<$	3.3E+12	&	2.6E+12	$-$	2.6E+13		&														$>$	0.8 	\\
G224S	&	9.6E+12			&	1.4E+13	$-$	1.4E+14		&										1.5 	$-$	15 				\\
G224NE	&	8.6E+12			&	7.7E+12				&				0.9 						\\
G224NW	&	1.2E+13			&	6.4E+11	$-$	6.4E+12		&			0.05 	$-$	0.5 				\\
\enddata
\end{deluxetable}

\clearpage

\floattable
\begin{deluxetable}{lcccccccccc}
\rotate
\tablecaption{Column Densities and Their Ratios toward T70 or H22 Position \label{tbl-9}}
\tablenum{9}
\tablecolumns{11}
\tablewidth{0pt}
\tablehead{
\colhead{Source} &
\colhead{$N$(CCS)}   &
\colhead{$N$(N$_2$H$^+$)} &
\colhead{$N$(N$_2$D$^+$)}   &
\colhead{$N$(HN$^{13}$C)}  &
\colhead{$N$(DNC)}  &
\colhead{$N$(NH$_3$)} &
\colhead{$N$(N$_2$H$^+$)/$N$(CCS)}   &
\colhead{$N$(DNC)/$N$(HN$^{13}$C)}   &
\colhead{$N$(N$_2$D$^+$)/$N$(N$_2$H$^+$)}   &
\colhead{$N$(NH$_3$)/$N$(CCS)}\\
\colhead{} &
\colhead{(cm$^{-2}$)}   &
\colhead{(cm$^{-2}$)} &
\colhead{(cm$^{-2}$)} &
\colhead{(cm$^{-2}$)}   &
\colhead{(cm$^{-2}$)} &
\colhead{(cm$^{-2}$)} &
\colhead{} &
\colhead{} &
\colhead{} &
\colhead{}
}
\startdata
G089.9-01.9	&	6.5E+12			&	4.2E+12				&	5.9E+12	&		1.5E+12	&	7.6E+12	&	5.2E+14	$\pm$	9.5E+13			&	0.7 								&		5.0 						&	1.4 	&	80 	$\pm$	15 					\\
G108N	&	4.8E+12			&					&		&	$<$	7.5E+11	&	1.9E+12	&	5.1E+14	$\pm$	1.7E+14			&									&			$>$	2.5 				&		&	106 	$\pm$	36 					\\
G108S	&				&					&		&			&		&				$<$	2.6E+14	&									&								&		&								\\
G120N	&	3.8E+12			&	8.9E+12				&		&		1.2E+12	&	2.0E+12	&	4.0E+14	$\pm$	2.0E+14			&	2.4 								&		1.7 						&		&	106 	$\pm$	53 					\\
G120S 	&	7.0E+12			&					&		&			&		&	5.2E+14	$\pm$	1.4E+14			&									&								&		&	74 	$\pm$	19 					\\
G149.5-1.2	&		$<$	3.0E+12	&	7.2E+12				&		&		2.3E+11	&	7.4E+11	&				$<$	1.9E+14	&					$>$	2.4 			&		3.2 						&		&								\\
G157.6-12.2	&	5.7E+12			&	1.4E+13				&	5.7E+12	&		1.3E+12	&	1.4E+13	&				$<$	2.4E+14	&	2.4 								&		10.4 						&	0.4 	&				$<$		42 		\\
G174.0-15.8	&	1.1E+13			&		$<$	7.0E+12		&		&			&		&				$<$	1.9E+14	&		$<$	0.6 						&								&		&				$<$		17 		\\
G192.33-11.88	&		$<$	4.8E+12	&	1.5E+13				&	1.6E+12	&		7.8E+11	&	4.8E+12	&	3.4E+14	$\pm$	9.2E+13			&					$>$	3.1 			&		6.2 						&	0.1 	&				$>$			71 	\\
G202.00+2.65	&				&					&		&			&		&				$<$	5.1E+14	&									&								&		&								\\
G202.31-8.92	&				&					&		&			&		&				$<$	2.6E+14	&									&								&		&								\\
G204.4-11.3	&	1.2E+13			&	2.0E+13				&	4.3E+12	&		3.5E+12	&		&	4.6E+14	$\pm$	7.8E+13			&	1.6 								&								&	0.2 	&	37 	$\pm$	6 					\\
G207N\tablenotemark{b}	&		$<$	6.3E+12	&					&		&			&		&	2.1E+14	$\pm$	1.5E+14			&									&								&		&								\\
G207S	&				&					&		&			&		&	2.3E+14	$\pm$	2.1E+14			&									&								&		&								\\
G224.4-0.6	&		$<$	2.7E+12	&	7.7E+12				&		&			&		&						&					$>$	2.9 			&								&		&								\\
\enddata
\tablenotetext{b}{incorrect position}
\end{deluxetable}

\clearpage

\begin{deluxetable}{lcccc}
\tablenum{10}
\tabletypesize{\scriptsize} \tablecolumns{5} 
\tablewidth{0pc}
\tablecaption{Fractional abundance toward the JCMT/SCUBA-2 Peak} \tablehead{
\colhead{Source} &
\colhead{$S_{\nu}^{beam}$\tablenotemark{c} } &
\colhead{$N_{H_{2}}$} &
\colhead{$X$(CCS)}&
\colhead{$X$(N$_2$H$^+$)}\\
\colhead{} &
\colhead{(mJy/beam)} &
\colhead{(10$^{22}$ cm$^{-2}$)} &
\colhead{} &
\colhead{} 
}
\startdata
G089E	&	132.2 	&	1.7 	&	2.0E-10	&	7.3E-10	\\
G089W	&	235.7 	&	3.0 	&	2.1E-10	&	1.5E-10	\\
G108N	&	231.9 	&	1.8 	&	2.9E-10	&	3.0E-10	\\
G108S1	&	167.3 	&	0.8 	&	8.0E-10	&		\\
G108S2	&	197.5 	&	1.0 	&	1.1E-09	&	2.5E-09	\\
G108S3	&	175.9 	&	0.9 	&		&		\\
G108S4	&	141.7 	&	0.7 	&		&	1.8E-09	\\
G120S1	&	332.1 	&	3.5 	&	2.8E-10	&	3.6E-10	\\
G120S2	&	116.9 	&	1.2 	&		&	1.7E-09	\\
G120N	&	514.4 	&	4.1 	&	9.3E-11	&	8.7E-11	\\
G149N	&	117.2 	&	0.9 	&		&	8.0E-10	\\
G149S	&	93.2 	&	0.7 	&		&		\\
G157	&	126.8 	&	1.5 	&	5.9E-10	&	1.7E-09	\\
G174	&	90.3 	&	1.1 	&	4.5E-10	&		\\
G192N	&	600.3 	&	6.0 	&		&		\\
G192S	&	333.3 	&	3.3 	&		&	5.0E-10	\\
G204NE	&	861.5 	&	12.8	&	9.5E-11	&	1.3E-10	\\
G204SW	&	338.5 	&	5.0 	&		&	1.3E-10	\\
G207N1	&	386.4 	&	3.1	&		&	4.0E-10	\\
G207N2	&	216.2 	&	1.7	&		&	4.5E-10	\\
G207N3	&	185.3 	&	1.5 	&		&		\\
G207N4	&	163.7 	&	1.3 	&		&	5.2E-09	\\
G207N5	&	121.9 	&	1.0 	&		&		\\
G224S	&	940.3	&	6.5 	&	1.5E-10	&		\\
G224NE	&	946.5	&	6.5 	&	1.3E-10	&	1.2E-10	\\
G224NW	&	387.5 	&	2.7 	&	4.4E-10	&		\\
\hline										
Median	&	206.9 	&	1.7 	&	2.8E-10	&	4.5E-10	\\
\enddata
\tablenotetext{c}{Convolved with a beam of 18.8 arcsec}
\end{deluxetable}

\clearpage

\begin{deluxetable}{lccccccc}
\tablenum{11}
\tabletypesize{\scriptsize} \tablecolumns{8} 
\tablewidth{0pc}
\tablecaption{Fractional abundance toward the T70 Position} \tablehead{
\colhead{Source} &
\colhead{$S_{\nu}^{beam}$\tablenotemark{c} } &
\colhead{$N_{H_{2}}$ } &
\colhead{$X$(CCS)}&
\colhead{$X$(N$_2$H$^+$)}&
\colhead{$X$(N$_2$D$^+$)}&
\colhead{$X$(HN$^{13}$C)}&
\colhead{$X$(DNC)}\\
                 &
\colhead{(mJy/beam)} &
\colhead{(10$^{22}$ cm$^{-2}$)} &
&
&
&
&
}
\startdata
G089.9-01.9	&	277.7 	&	2.0 	&	3.3E-10	&	2.1E-10	&	2.9E-10	&	7.5E-11	&	3.8E-10	\\
G108N	&	225.4 	&	1.0 	&	4.8E-10	&		&		&		&	1.9E-10	\\
G120N	&	629.3 	&	2.8 	&	1.3E-10	&	3.2E-10	&		&	4.3E-11	&	7.2E-11	\\
G149.5-1.2	&	108.2 	&	0.4 	&		&		&		&	5.9E-11	&	1.9E-10	\\
G157.6-12.2	&	162.3 	&	1.1 	&	5.2E-10	&	1.2E-09	&	5.2E-10	&	1.2E-10	&	1.2E-09	\\
G192.33-11.88	&	623.9 	&	3.5 	&		&	4.2E-10	&	4.7E-11	&	2.2E-11	&	1.4E-10	\\
G204.4-11.3	&	1128.8 	&	9.3 	&	1.3E-10	&	2.1E-10	&	4.6E-11	&	3.8E-11	&		\\
G224.4-0.6	&	88.4 	&	0.3 	&		&		&		&		&		\\
\hline															\\
Median	&	251.6 	&	1.6 	&	3.3E-10	&	3.2E-10	&	1.7E-10	&	5.1E-11	&	1.9E-10	\\
\enddata
\tablenotetext{c}{Convolved with a beam of 18.8 arcsec}
\end{deluxetable}

\clearpage

\begin{deluxetable}{lc}
\tablenum{12}
\tabletypesize{\scriptsize} \tablecolumns{2} \tablewidth{0pc}
\tablecaption{CEF  based on the $N$(N$_2$H$^+$)/$N$(CCS) column density ratio toward the starless  SCUBA-2 Peak} \tablehead{
\colhead{SCUBA-2 peak} &
\colhead{CEF(N$_2$H$^+$,CCS)} 
}
\startdata
G089E	&	(	8 					)	\\
G089W	&	(	-27 					)	\\
G108N	&	(	-19 					)	\\
G108S2	&	(	-2 					)	\\
G120S2	&	(				$>$	19 	)	\\
G120N	&	(	-21 					)	\\
G157	&		3 						\\
G174	&		-44 	$-$	6 				\\
G204NE	&		-13 						\\
G207N2	&					$>$	-2 		\\
G207N4	&					$>$	46 		\\
G207N5	&					$>$	-25 		\\
G224NW	&	(	-83 	$-$	-33 			)	\\
\enddata
\end{deluxetable}
\clearpage

\begin{deluxetable}{lcccc}
\tablenum{13}
\tabletypesize{\scriptsize} \tablecolumns{5} \tablewidth{0pc}
\tablecaption{CEF toward the starless T70 Position} \tablehead{
\colhead{Source} &
\colhead{CEF(N$_2$H$^+$,CCS) } &
\colhead{CEF(DNC,HN$^{13}$C) } &
\colhead{CEF(N$_2$D$^+$,N$_2$H$^+$)}&
\colhead{CEF(average)}
}
\startdata
G089.9-01.9	&	(			-29 	)	&	(			27 	)	&	(	33 	)	&	(	10 	$\pm$	34 	)	\\
G108N	&						&	(	-9 			)	&				&	(	-9 			)	\\
G120N	&	(			-1 	)	&	(			-30 	)	&				&	(	-16 	$\pm$	20 	)	\\
G157.6-12.2	&				-1 		&				65 		&		7 		&		24 	$\pm$	36 		\\
G174.0-15.8	&		$<$	-30 			&						&				&						\\
G204.4-11.3	&				-10 		&						&		-7 		&		-8 	$\pm$	2 		\\
G224.4-0.6	&	(	$>$	3 		)	&						&				&						\\
\enddata
\end{deluxetable}

\clearpage

\acknowledgments

M. K. was supported by Basic Science Research Program through the National Research Foundation of Korea(NRF) funded by the Ministry of  Science, ICT \& Future Planning (No. NRF-2015R1C1A1A01052160).
M. J. acknowledges the support of the Academy of Finland Grant No. 285769.
The James Clerk Maxwell Telescope is operated by the East Asian Observatory 
on behalf of The National Astronomical Observatory of Japan, 
Academia Sinica Institute of Astronomy and Astrophysics, 
the Korea Astronomy and Space Science Institute, 
the National Astronomical Observatories of China and the Chinese Academy of Sciences (Grant No. XDB09000000), 
with additional funding support from the Science and Technology Facilities Council of the United Kingdom and participating universities in the United Kingdom and Canada.



\vspace{5mm}
\facilities{JCMT, No:45m}

\software{AIPS, NewStar, NOSTAR}

\end{document}